\documentclass[%
 aip,
 amsmath,amssymb,
 reprint 
]{revtex4-2}

\usepackage{graphicx}
\usepackage{dcolumn}
\usepackage{bm}
\usepackage[utf8]{inputenc}
\usepackage[T1]{fontenc}
\usepackage{mathptmx}
\usepackage[hidelinks]{hyperref}
\usepackage{adjustbox}
\usepackage{indentfirst}
\usepackage{xcolor}

\usepackage{tikz}
\usetikzlibrary{arrows.meta,positioning}
\AtBeginDocument{%
  \setlength{\parindent}{2em}
  \setlength{\parskip}{0pt}
}

\begin{document}

\title{Patient-Scale Blood Flow Analysis in Artery Stent Implantation via Smoothed-Particle Hydrodynamics}

\author{Jinlei Zhou}
\affiliation{Chair of Aerodynamics and Fluid Mechanics, Technical University of Munich, 85748 Munich, Germany}

\author{Sukang Peng}
\affiliation{Chair of Aerodynamics and Fluid Mechanics, Technical University of Munich, 85748 Munich, Germany}

\author{Yongchuan Yu}
\affiliation{Chair of Space Propulsion and Mobility, Technical University of Munich, 85521 Ottobrunn, Germany}

\author{Dong Wu}
\email[Author to whom correspondence should be addressed: ]{dong.wu@tum.de}
\affiliation{Chair of Aerodynamics and Fluid Mechanics, Technical University of Munich, 85748 Munich, Germany}

\author{Xiangyu Hu}
\affiliation{Chair of Aerodynamics and Fluid Mechanics, Technical University of Munich, 85748 Munich, Germany}

\date{Submitted: 11 November 2025}

\begin{abstract}
A unified Smoothed Particle Hydrodynamics (SPH) simulation framework for coronary stent implantation is developed, 
which unifies weakly-compressible hemodynamics, Neo-Hookean solids, and stent–artery contacts, based on a multi-resolution particle discretization. 
Prior to application, feasibility and accuracy are established via three baseline validations: 
(i) poiseuille flow in a two-dimensional channel with prescribed parabolic inflow and a pressure outlet, maintaining parabolic profiles with low Root Mean Squared Error of Prediction (RMSEP);
(ii) channel flow initialized with a uniform velocity field and driven by a specified inlet–outlet pressure differential, with agreement to reference profiles quantified by low RMSEP at five reference instants;
and (iii) a three-ring impact benchmark in solid mechanics, capturing large deformation, multi-body contact, and self-contact.
The validated framework is subsequently applied to a coronary bifurcation with a focal stenosis, 
where flow-field diagnostics reveal acceleration at the stenotic throat, near-wall low-velocity zones, and co-localization of elevated pressure with increased Von-Mises stress at the bifurcation and inlet. 
Following simulated stent implantation, velocity transitions across the stenosis become smoother, pressure gradients are reduced, and the fractional flow reserve increases from 0.45 to 0.91. 
These results demonstrate that the proposed SPH framework yields quantitatively reliable, clinically interpretable hemodynamic metrics alongside robust solid–solid contact predictions, thereby supporting rigorous analysis and pre-procedural planning of vascular interventions.
\end{abstract}

\maketitle

\section{Introduction}
\begin{figure}
    \centering
    \includegraphics[width=\linewidth]{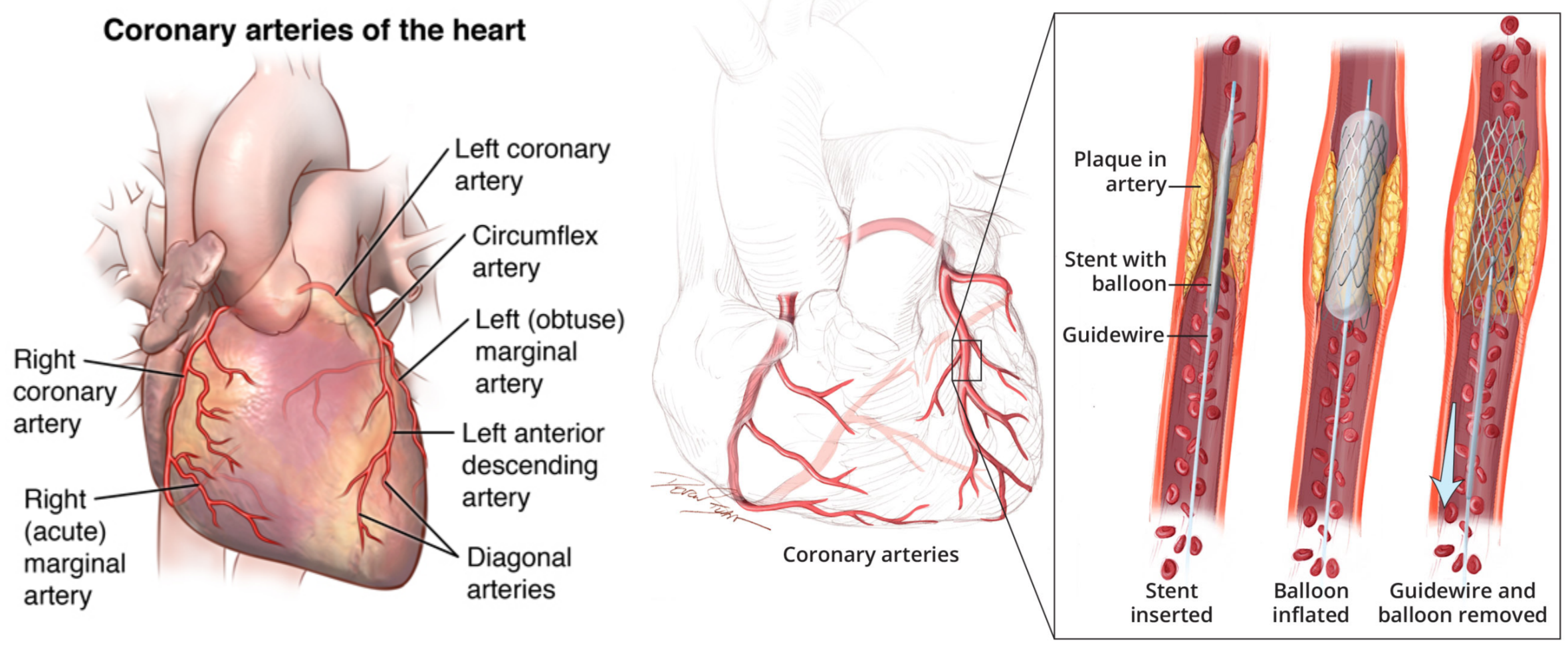}
    \caption{Left: anatomical locations of the right and left coronary arteries\,\cite{TexasHeartInstitute}. Right: steps of percutaneous stent implantation from guidewire passage and balloon inflation to final stent deployment \,\cite{nhlbi_stent_placement}.}
    \label{fig:1}
\end{figure}

Coronary arteries are elastic tubes that carry oxygen-rich blood from the heart to the heart muscle and other organs, smoothing the pulses of each heartbeat \cite{Braunwald2022, Sunthareswaran2022, GraysAnatomy2016}. 
The coronary circulation arises from the ascending aorta and runs over the cardiac surface; the left coronary artery typically bifurcates into the left anterior descending (LAD) and circumflex (LCx) branches (\textcolor{blue}{\hyperref[fig:1]{Fig.~1}}, left), which perfuse the anterior and lateral walls of the left ventricle, respectively. 
The compliant arterial wall accommodates pulsatile pressure and adjusts flow to match myocardial metabolic demand, thereby sustaining normal cardiac function \cite{Klabunde2012}. 
Coronary artery disease (CAD) happens when atherosclerotic plaque builds up and is one of the most common vascular diseases\cite{Ross1999}. This narrows the lumen and slows blood flow, raising the risk of events like heart attacks \cite{RobbinsCotran2014,Libby2002,Gutierrez2022,Holmstedt2013}. Treatment includes controlling risk factors and using medicines (lowering lipids, managing blood pressure and glucose, and antiplatelet therapy) \cite{Haller2002}. When needed, doctors restore blood flow by surgery or procedures such as endarterectomy, bypass, or Peripheral Component Interconnect (PCI). Balloon angioplasty with a stent is the main way to reopen the vessel \cite{Kandarpa2001}. Coronary stents are small metal lattice tubes (typically stainless steel or nitinol) placed by PCI ; balloon inflation expands the device to re-establish patency (\textcolor{blue}{\hyperref[fig:1]{Fig.~1}}, right). Contemporary devices include bare-metal stents (BMS) and drug-eluting stents (DES), the latter releasing antiproliferative agents to lower re-stenosis risk. Stenting alleviates angina, improves perfusion, and reduces adverse cardiac events \cite{Topol1998}.

Good coronary hemodynamics are key to organ perfusion \cite{Rutherford2010,Brown1994}, but in difficult cases decisions still rely largely on the operator’s experience\cite{Xu2016LeftMainPCIExperience}. Therefore, computational hemodynamic simulation has become a useful complement to imaging and clinical judgment \cite{Perktold1995}. Predictive simulations can help: they test whether a plan is feasible, suggest the stent type, size, and placement, and estimate potential benefits such as pressure recovery and fractional flow reserve (FFR), which is especially useful in curved or bifurcating lesions \cite{pijls1996measurement}.

Prior works into coronary stent implantation reveal both capability and gaps, spanning two complementary strands: structural mechanics and hemodynamics.
On the structural side, the Finite Element Method (FEM) is most common. It has shown how stent design and size affect contact mechanics and wall loading during expansion, resolved patient-specific stress and strain patterns, and shown that physiologic dynamic bending increases alternating stress and localizes fatigue-prone regions. It also provides credible post-implant lumens for downstream flow analysis. These studies treat the solid problem alone, modeling stent shape and stresses without the blood flow \cite{Takashima2007, Gijsen2008, Zhao2016, Zhao2021}.
On the flow side, 
three-dimensional Computational Fluid Dynamics (CFD) uses either the traditional Finite Volume Method (FVM) or the newer Smoothed Particle Hydrodynamics (SPH) to model the fluid. 
With FVM, stent geometry has been shown to reorganize near-wall velocity and wall shear stress (WSS) from the proximal to distal segments \cite{LaDisa2003}. 
Optical coherence tomography (OCT) – based, 
patient-specific pipelines also support transient simulations with physiologic inflow and branch partitioning  while the flow is solved with rigid vessel walls in these studies, which suits low curvature and low Reynolds number and therefore is not ideal for curved, 
diseased coronary arteries and arterial flow \cite{Migliori2020}. 
With SPH, in vascular applications: it represents blood with particles and treats the vessel wall with boundary particles, allowing direct evaluation of near-wall velocity, WSS but the vessel is also modeled as rigid wall \cite{Napoli2019}. 
There are also studies that improve driving low-Reynolds-number flow and handling curved walls while still limited to low curvature and low Reynolds number flow \cite{Liang2012, Monteleone2019AneurysmSPH}. 
When vessels are deformable, SPH can achieve patient-specific fluid–structure interaction with Windkessel outlets but many comparisons still rely on non-bifurcating or simplified geometries \cite{Lu2024}.

Despite this progress, truly end-to-end frameworks that co-simulate structure and flow throughout bifurcation stenting are still uncommon. 
To address the limitation, SPH offers a mesh-free, total-Lagrangian alternative in which particles move with the material, naturally accommodating complex coronary geometries, large deformations, and moving interfaces while avoiding mesh distortion \cite{Liu2005SPH,Ye2019, Violeau2006, XiangChen2015}. 
Since many common physical abstractions can be expressed as particle interactions, SPH can discretize multi-physics within a unified particle framework, allowing core algorithms such as neighbor search, time stepping, and data handling to be shared, which simplifies parallelization and improves efficiency \cite{Zhang2021SPHinXsys, Sun2021FSISPH}. 
Just as importantly, such a unified framework supports strongly coupled formulations \cite{Matthies2003PartitionedStrongCoupling,Matthies2006StrongCouplingProcedures}. Moreover,  reviews report engineering-level accuracy with strong potential for GPU-accelerated use\cite{mokhtar2015review}. These advantages motivate our use of SPH to integrate prescribed blood flow, stent mechanics, and vessel deformation within one particle-based framework.

In this work, a unified SPH simulation framework for coronary stent implantation is developed, which unifies incompressible hemodynamics and stent–artery contacts, based on a multi-resolution particle discretization. Concretely, we (i) run simulations with Windkessel outlets with coronary geometries in pre- and post-stent implantations ; and (ii) quantify outcomes via pressure drop, WSS maps, flow splitting, and FFR. By avoiding meshing bottlenecks, the approach aims to lessen reliance on operator experience and provide quantitative guidance for complex lesions, including choosing stent type and size and anticipating post-stent flow behavior.

The rest of the paper is organized as follows: Section II explains the numerical method (Lagrangian conservation laws, Weakly Compressible Smoothed Particle Hydrodynamics (WCSPH) with an isothermal Equation of State, Riemann fluxes, wall boundaries, total-Lagrangian solid, and dual-criteria time stepping with Verlet). Section III validates the solver (Poiseuille Channel Flow with parabolic profiles ; a pressure-driven channel benchmark at five time points); and a three-ring impact with large deformation . Section IV applies the framework to a patient-oriented coronary bifurcation with a focal stenosis, simulates stent expansion, and compares pre-flow and post-flow results, pressure, wall stress, and FFR. Section V concludes and outlines steps toward a full patient-specific pipeline. For transparency and future work, our codes are available in the SPHinXsys open-source repository \cite{Zhang2021multiresolution,SPHinXsysGithub}: \url{https://github.com/Xiangyu-Hu/SPHinXsys}.

\section{Numerical Method}
\subsection{Governing Equations of Fluid Dynamics}
Within a Lagrangian description, the conservation laws of mass and momentum are written as
\begin{equation} \left\{ \begin{aligned} \frac{d\rho}{dt} &= -\rho \nabla \cdot \mathbf{v}, \\[8pt]\frac{d\mathbf{v}}{dt} &= -\frac{1}{\rho}\nabla p + \nu\nabla^2 \mathbf{v}. \end{aligned} \right. \end{equation}
In these expressions, $\rho$ denotes fluid density, $\mathbf{v}$ the velocity field, $p$ the pressure, $\nu$ the kinematic viscosity. The material (Lagrangian) derivatives $d\rho/dt$ and $d\mathbf{v}/dt$ represent temporal variations experienced by fluid particles as they move with the flow. The mass equation states that density changes arise from the divergence of velocity; for incompressible motion, $d\rho/dt=0$, which reduces to the solenoidal constraint $\nabla\cdot\mathbf{v}=0$. The momentum equation balances unsteady acceleration with pressure-gradient, viscous, and body-force contributions. In the Weakly Compressible Smoothed Particle Hydrodynamics (WCSPH) formulation, pressure is further related to density through an isothermal artificial equation of state \cite{Morris1997}:
\begin{equation}
p = c_0^2 \left( \rho - \rho_0 \right).
\end{equation}
Here $c_0$ is a prescribed acoustic speed, commonly selected as $c_0 = 10*U_{\max}$, where $U_{\max}$ denotes the anticipated maximum flow speed \cite{Monaghan1994}.

\subsection{WCSPH Method Based on Riemann Solvers}
A Riemann solver provides a local interfacial solution of the conservation laws and thus reliable numerical fluxes across discontinuities. In WCSPH, these fluxes are used to improve stability and accuracy of particle interactions \cite{toro2013riemann}. The semi-discrete continuity and momentum equations \cite{zhang2017weakly} are
\begin{equation}
\left\{
\begin{aligned}
\frac{d\rho_i}{dt} &= 2\rho_i \sum_j \frac{m_j}{\rho_j}\,(U^*-\mathbf{v}_i)\!\cdot\!\mathbf{e}_{ij}\,
\frac{dW_{ij}}{dr_{ij}},\\[6pt]
\frac{d\mathbf{v}_i}{dt} &= -\,m_i \sum_j \frac{2P_j^*}{\rho_i\rho_j}\,\nabla_i W_{ij}
\;+\; m_i \sum_j \frac{2\mu}{\rho_i\rho_j}\,\frac{\mathbf{v}_{ij}}{r_{ij}}\,
\frac{dW_{ij}}{dr_{ij}}\, ,
\end{aligned}
\right.
\end{equation}
here, $m_j$ and $\rho_j$ are the mass and density of neighbor $j$, $U^*$ is the interface-normal velocity from the Riemann solution, $\mathbf{e}_{ij}$ is the unit vector along the $i$–$j$ line, $W_{ij}$ is the kernel, and $r_{ij}$ is the inter-particle distance. The momentum update combines the interfacial pressure $P_j^*$ and a viscous term with dynamic viscosity $\mu$; $\nabla_i W_{ij}$ is the kernel gradient and $\mathbf{v}_{ij}=\mathbf{v}_i-\mathbf{v}_j$ is the relative velocity.

The left and right states for the inter-particle Riemann problem are
\begin{equation}
\left\{
\begin{aligned}
(\rho_L, U_L, P_L, c_L)\ &= (\rho_i, -\mathbf{v}_i\cdot \mathbf{e}_{ij}, p_i, c_i), \\[5pt]
(\rho_R, U_R, P_R, c_R)\ &= (\rho_j, -\mathbf{v}_j\cdot \mathbf{e}_{ij}, p_j, c_j).
\end{aligned}
\right.
\end{equation}
Velocities are projected onto $\mathbf{e}_{ij}$ so the problem is posed along the interaction line. With a piecewise-constant reconstruction, the solution has two outer waves (shock or rarefaction) and a contact discontinuity that separates the intermediate states $(\rho_L^*,U^*_L,P^*_L)$ and $(\rho_R^*,U^*_R,P^*_R)$. Imposing
\begin{equation}
U_L^* = U_R^* = U^*, 
\quad P_L^* = P_R^* = P^*,
\end{equation}
yields a unique $(U^*,P^*)$ and ensures consistent transfer of momentum and pressure.

\begin{figure}
  \centering
  \includegraphics[width=1.0\linewidth]{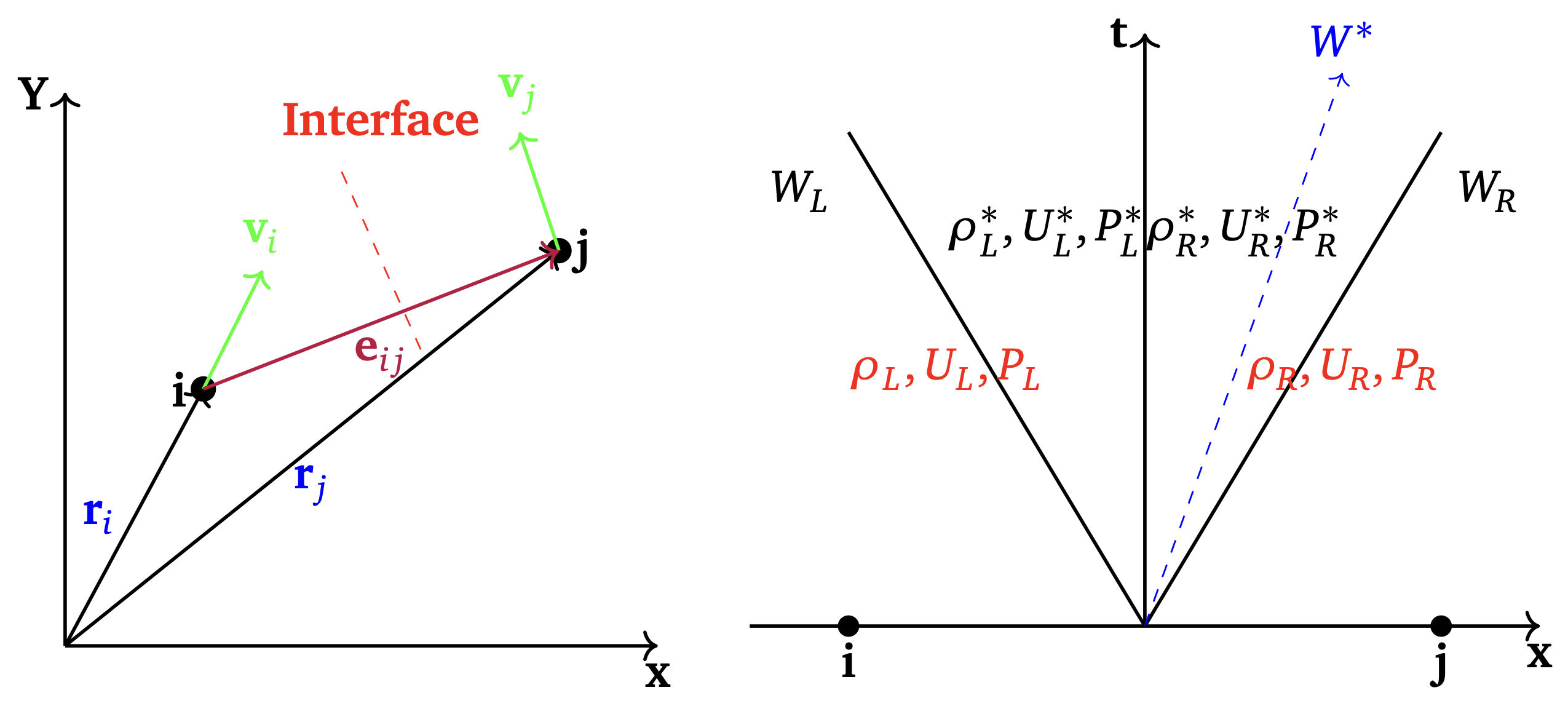}
  \caption{Inter-particle Riemann problem: construction along the $i$–$j$ line (left) and wave structure of the solution (right) \cite{zhang2017weakly}.}
  \label{fig:2}
\end{figure}

Assuming an identical sound speed in both states, the star state is approximated by
\begin{equation}
\left\{
\begin{aligned}
U^* &= \overline{U} + \frac{P_L - P_R}{c(\rho_L + \rho_R)}, \\[6pt]
P^* &= \overline{P} + \frac{\rho_L \rho_R \beta (U_L - U_R)}{\rho_L + \rho_R}.
\end{aligned}
\right.
\end{equation}
These closed forms provide efficient estimates of the interfacial velocity $U^*$ and pressure $P^*$ for WCSPH flux calculations in variable-density or multi-phase flows (\textcolor{blue}{\hyperref[fig:2]{Fig.~2}}).

\subsection{Wall Boundary Conditions}
\label{subsec:wall-boundaries}

\begin{figure}[t]
  \centering
  \includegraphics[width=0.6\linewidth]{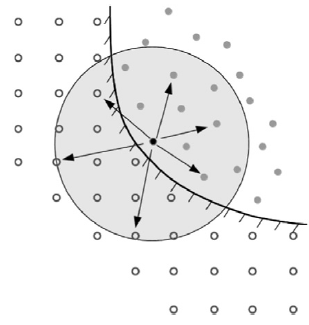}
  \caption{Wall boundary particles: fluid particles (solid dots) interact with Cartesian dummy particles (hollow) to complete kernel support and impose wall conditions \cite{adami2012generalized}.}
  \label{fig:3}
\end{figure}

For walls, a pressure boundary condition based on local force balance prevents particle penetration and applies to both stationary and moving walls \cite{adami2012generalized}. For complex geometries, dummy (ghost) particles approximate the wall surface; these particles contribute to the updates of continuity and momentum via kernel overlap. The continuity equation assigns initial particle volume to wall particles and prescribes the wall velocity $\mathbf{v}_w$. As fluid particles approach the wall, density growth induces a pressure force that enforces impermeability ($\mathbf{v}\!\cdot\!\mathbf{n}=0$). A free-slip wall neglects viscous terms at dummy particles; a no-slip wall extrapolates a smoothed velocity to the dummy locations (\textcolor{blue}{\hyperref[fig:3]{Fig.~3}}):
\begin{equation}
\widetilde{\mathbf{v}}_a = \frac{\sum_b \mathbf{v}_b\, W_{ab}}{\sum_b W_{ab}}.
\end{equation}
The dummy velocity is then taken as $\mathbf{v}_w = 2\mathbf{v}_s - \widehat{\mathbf{v}}_a$, where $\mathbf{v}_s$ is the prescribed wall speed. Pressure on wall particles is obtained from a fluid–wall force balance, accounting for wall motion\cite{marrone2010fast},
\begin{equation}
\frac{d \mathbf{v}_f}{dt} = -\frac{\nabla p_f}{\rho_f} + \mathbf{g} = \mathbf{a}_w,
\end{equation}
which yields
\begin{equation}
\int \nabla p \cdot d\mathbf{l} = \rho_f \int \bigl(\mathbf{g} - \mathbf{a}_w\bigr)\cdot d\mathbf{l},
\quad
p_w = p_f + \rho_f \bigl(\mathbf{g} - \mathbf{a}_w\bigr)\cdot \mathbf{r}_{wf}.
\end{equation}
Summing contributions from neighboring fluid particles (excluding dummy ones) gives the wall density–pressure relation
\begin{equation}
\rho_w = \rho_{0,b}\left( \frac{p_w - \chi}{p_{0,b}} + 1 \right)^{\frac{1}{\gamma}}.
\end{equation}

\subsection{Total-Lagrangian Solid Formulation}
\label{subsec:tl-sph}

For solids, the Total-Lagrangian SPH (TLSPH) scheme is adopted: neighbors are defined in the reference configuration and remain fixed; mass and momentum conservation are discretized as
\begin{equation}
\left\{
\begin{aligned}
\rho_i &= \rho^0 \,\frac{1}{\det(\mathbf{F})}, \\[8pt]
\frac{d\mathbf{v}_i}{dt} &= \frac{2}{m_i}\sum_j V_i V_j \,\widetilde{\mathbf{P}}_{ij}\,\nabla_i^0 W_{ij} + \mathbf{g}.
\end{aligned}
\right.
\end{equation}
The symmetric Piola–Kirchhoff stress for a particle pair $(i,j)$ is\cite{Zhang2021multiresolution}
\begin{equation}
\widetilde{\mathbf{P}}_{ij} = \frac{1}{2}\bigl(\mathbf{P}_i\,\mathbf{B}_i^0 + \mathbf{P}_j\,\mathbf{B}_j^0\bigr),
\end{equation}
where the correction matrix \cite{Vignjevic2006} is
\begin{equation}
\mathbf{B}_i^0 = \left(\sum_j V_j\,\bigl(\mathbf{r}_j^0 - \mathbf{r}_i^0\bigr)\otimes \nabla_i^0 W_{ij}\right)^{-1}.
\end{equation}
This matrix compensates for errors due to non-uniform reference particle layouts and improves the consistency of the deformation-gradient approximation.

Adhering to the Courant–Friedrichs–Lewy (CFL) condition and the limit under external force ~\cite{Hirsch1988,Liu2003SPH}, the time step size is given as
\begin{equation}
\Delta t \;=\; \mathrm{CFL}\;
\min\!\left(
\frac{h}{\,C + |\dot{\mathbf{v}}|_{\max}\,},\;
\sqrt{\frac{h}{\,|\ddot{\mathbf{v}}|_{\max}\,}}
\right),
\tag{12}
\end{equation}
where the CFL number is set to $0.6$, as recommended in Refs.~\cite{Zhang2021multiresolution}.

\subsection{Dual-Criteria Fluid Time Stepping}

Dual-criteria time stepping improves efficiency for weakly compressible SPH by separating the
advection and acoustic constraints \cite{Zhang2020DualCriteria}. A larger advection step moves
particles with the flow, while several smaller acoustic (pressure–relaxation) substeps advance
the fast compressible dynamics. This reduces interaction updates and lowers the overall cost.

\begin{equation}
\Delta t_{\text{ad}} = CFL_{\text{ad}} \min\!\left(\frac{h}{\lvert \mathbf{v}\rvert_{\max}}, \frac{h^2}{\nu}\right),
\end{equation}
 $\Delta t_{\text{ad}}$ is the advection time step, $CFL_{\text{ad}}{=}0.25$ is the Courant number,
$h$ is the smoothing length, $\lvert \mathbf{v}\rvert_{\max}$ is the maximum particle speed, and
$\nu$ is the kinematic viscosity. The minimum enforces stability.

\begin{equation}
\Delta t_{\text{ac}} = CFL_{\text{ac}} \,\frac{h}{c + \lvert \mathbf{v}\rvert_{\max}},
\end{equation}

 $\Delta t_{\text{ad}}$ is the acoustic step, using $CFL_{\text{ac}}{=}0.6$, the artificial sound speed $c$, and the
same $h$ and $\lvert \mathbf{v}\rvert_{\max}$. Compared with standard WCSPH, this criterion admits larger
overall steps while maintaining stability.

Time integration adopts the Verlet scheme \cite{Verlet1967}. The velocity is first advanced to a
midpoint value; positions and density are then updated; finally, the velocity is corrected to the end of
the step.

\begin{table}
\caption{Update steps in the Verlet method.}
\label{tab:1}
\begin{ruledtabular}
\begin{tabular}{p{6cm}p{6cm}}
\textbf{Step} & \hspace*{-5em}\textbf{Update Equation} \\
\hline
1. Velocity midpoint update &
\hspace*{-5em}$\mathbf{v}_i^{n+\frac{1}{2}} = \mathbf{v}_i^{n} + \frac{1}{2}\,\Delta t_{\text{ac}} \left(\frac{d \mathbf{v}_i}{dt}\right)^n$ \\\\

2. Position and density update &
\hspace*{-6em}$\displaystyle
\left\{
\begin{aligned}
\mathbf{r}_i^{n+1} &= \mathbf{r}_i^{n} + \Delta t_{\text{ac}} \,\mathbf{v}_i^{n+\frac{1}{2}}, \\
\rho_i^{n+1} &= \rho_i^{n} + \frac{1}{2}\,\Delta t_{\text{ac}}\left(\frac{d \rho_i}{dt}\right)^{n+\frac{1}{2}}
\end{aligned}
\right.
$ \\\\

3. Final velocity update &
\hspace*{-5em}$\mathbf{v}_i^{n+1} = \mathbf{v}_i^{n} + \frac{1}{2}\,\Delta t_{\text{ac}}\left(\frac{d \mathbf{v}_i}{dt}\right)^{n+1}$ \\
\end{tabular}
\end{ruledtabular}
\end{table}

\section{Method Validation}

\subsection{Poiseuille Channel Flow}
For the first case, the imposed inlet velocity of a two-dimensional channel with gap $2d$ and streamwise length $L$ follows the classical start-up solution
\begin{equation}
\begin{aligned}
& u_x(y,t) = \frac{\Delta P}{2\,\eta L}\,y(d-y)+ \sum_{n=0}^{\infty} \frac{4\,\Delta P\,d^{2}}{\eta L \pi^3 (2n+1)^3}\
\\&*sin\!\left(\frac{\pi(2n+1)}{d}\,y\right)\exp\!\left(-\,\frac{(2n+1)^2 \pi^2 \eta}{\rho d^{2}}\,t\right) &,
\end{aligned}
\end{equation}

where $y\in[0,d]$ measures the distance from one wall (the full gap is $2d$), $\Delta P$ is the pressure drop over a length $L$ which is equal to 3.0, $\eta=\nu \rho$ is the dynamic viscosity, and $\rho$ the density.
The computational domain has $L=8.0$ and $d=1.0$; the outlet pressure is set to $0$.
Bidirectional buffers are used solely to insert particles at the inlet and remove them at the outlet.
To target a Reynolds number $Re=50$, the viscosity is chosen as
\begin{equation}
\eta = \nu \rho = \frac{\rho d^{2} \Delta P}{8 L Re},
\qquad \rho = 1000.
\end{equation}

Weakly compressible SPH is adopted with an artificial sound speed $c_0 = 100\,u_x^{\max}$, where
\begin{equation}
u_x^{\max} = \frac{d^{2}\Delta P}{8\,\eta L},
\end{equation}
to accurately capture the rapid initial transient. Fluid and wall particles are initialized on a Cartesian lattice with three uniform spacings
$dp=d/12$, $d/25$, and $d/50$.

\begin{figure}
  \centering
  \includegraphics[width=1.0\linewidth]{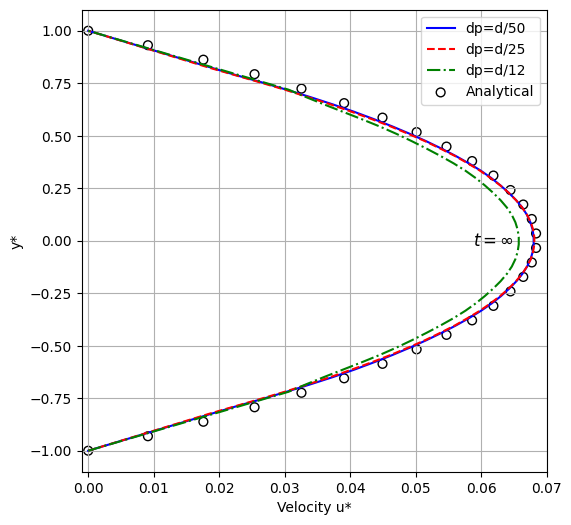}
  \caption{Comparison of numerical and analytical velocity profiles at different particle spacings.}
  \label{fig:4}
\end{figure}

Velocity profiles sampled at three different particle spacings  (\textcolor{blue}{\hyperref[fig:4]{Fig.~4}}) are compared with the analytical solution~(16).
Throughout the development, the numerical profiles remain parabolic, attain their maximum at the centerline, and approach zero at the plates under the no-slip boundary condition.
Following Holmes and Pivonka$^{8}$, we quantify accuracy by the root-mean-square error percentage (RMSEP) over the full profile,
\begin{equation}
\mathrm{RMSEP} \;=\;
\sqrt{\frac{1}{N}\sum_{i=1}^{N}
\left(\frac{u_x(y_i,t) - \tilde{u}_x(y_i,t)}{u_x(y_i,t)}\right)^{\!2}},
\end{equation}
where $N$ is the number of sampling points at time $t$, $u_x$ is the analytical velocity, and $\tilde{u}_x$ the numerical result.
The RMSEP is $6.31\%$ for $dp=d/12$, $3.87\%$ for $d/25$, and $3.40\%$ for $d/50$, indicating good agreement and clear convergence with refinement.

\subsection{Pressure-Driven Channel Flow}
We start the transient simulation with a fully developed velocity field, $u(y,$ $0)=1$, and add a constant pressure through the whole field.
The coordinate is non-dimensionalized by the half-height \(h\) and the velocity by \(U\), hence
\(y\in[-1,1]\) and \(u=u/U\).
To track the approach to the fully developed state, we define four reference instants \(t_k\) \((k=1,$ $2,$ $3,$ $4,$ $5)\)
as the times at which the maximum of the instantaneous velocity profile attains equally spaced levels
between the initial value \(1.0\) and the steady value \(1.5\), namely\begingroup
\setlength{\abovedisplayskip}{0.5pt}
\setlength{\belowdisplayskip}{0.5pt}
\begin{equation}
\max_{y\in[-1,1]} u(y,t_k) 
= 1.0 + (1.5-1.0)\frac{k}{5}
= 1.0 + 0.1k .
\end{equation}
\endgroup

The progressive rise of the peak velocity from \(1.0\) to \(1.5\) is accompanied by a corresponding growth of
the bulk flow rate during convergence.

\begin{table}[t]
\centering
\caption{RMSEP at the five reference instants.}
\label{tab:rmsep_fully_dev}
\begin{ruledtabular}
\begin{tabular}{@{}c@{\hspace{0.5em}}c@{}}
\hspace*{3em}\textbf{time steps $t$} & \hspace*{-30em}\textbf{RMSEP error (\%)}\\
\colrule
\hspace*{3em}$t_1$ & \hspace*{-35em}2.83\\
\hspace*{3em}$t_2$ & \hspace*{-35em}1.65\\
\hspace*{3em}$t_3$ & \hspace*{-35em}4.60\\
\hspace*{3em}$t_4$ & \hspace*{-35em}4.06\\
\hspace*{3em}$t_5$ & \hspace*{-35em}2.03\\
\end{tabular}
\end{ruledtabular}
\end{table}

\begin{figure}
  \centering
  \includegraphics[width=1.0\linewidth]{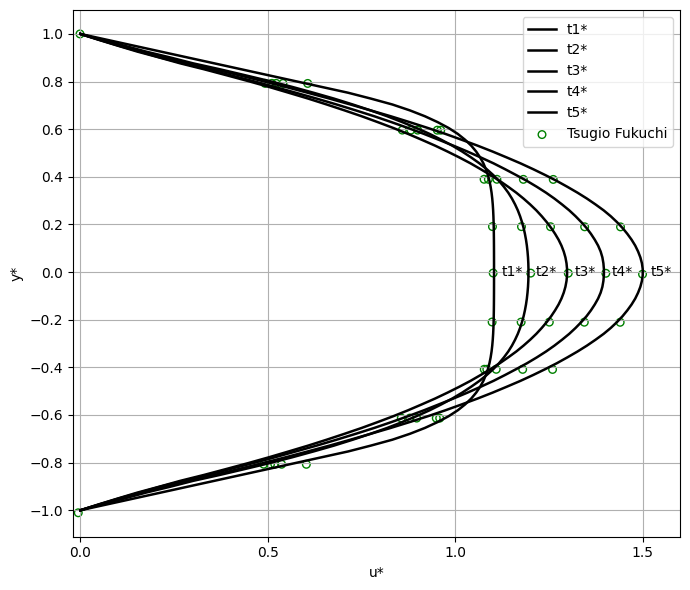}
  \caption{Velocity profiles \(u(y,t)\) at the five milestones \(t_k\). Symbols denote the reference data extracted from Tsugio Fukuchi\cite{Fukuchi2011},solid lines are the present SPH results.}
  \label{fig:5}
\end{figure}
The initial velocity is arbitrary. \textcolor{blue}{\hyperref[fig:5]{Fig.~5}} shows the calculation results setting the initial velocity at $u(y,$ $0)=1$. Each \(t_k\) \((k=1,$ $2,$ $3,$ $4,$ $5)\) is the time when the dimensionless velocity $u_5$ (the convergence maximum velocity, $u_{5\max} = 1.5$) reaches value $1.0 + (1.5 - 1.0) \times k/5$. The velocity profiles at \(t_k\) collapse with those reported by Tsugio Fukuchi\cite{Fukuchi2011} (green dots in \textcolor{blue}{\hyperref[fig:5]{Fig.~5}}). Quantitative comparison against the digitized reference points yields RMSEP below \(X\%\) for all stages \(k\) (\textcolor{blue}{\hyperref[tab:rmsep_fully_dev]{table II}}).

\begin{figure}
  \centering
  \includegraphics[width=1.0\linewidth]{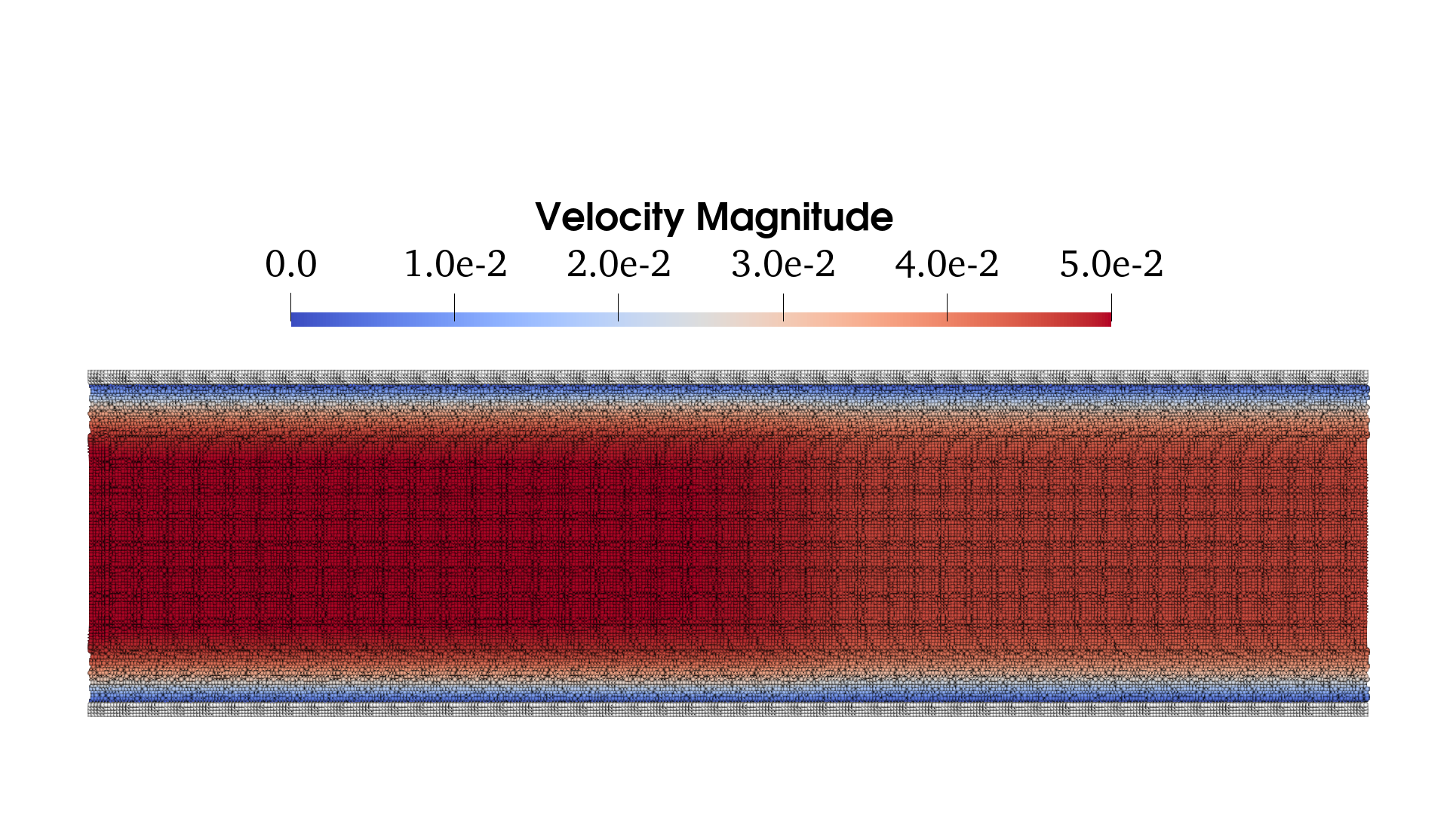}  
  \caption{\centering Velocity magnitude distribution at t1}
  \label{fig:6}
\end{figure}

\begin{figure}
  \centering
  \includegraphics[width=1.0\linewidth]{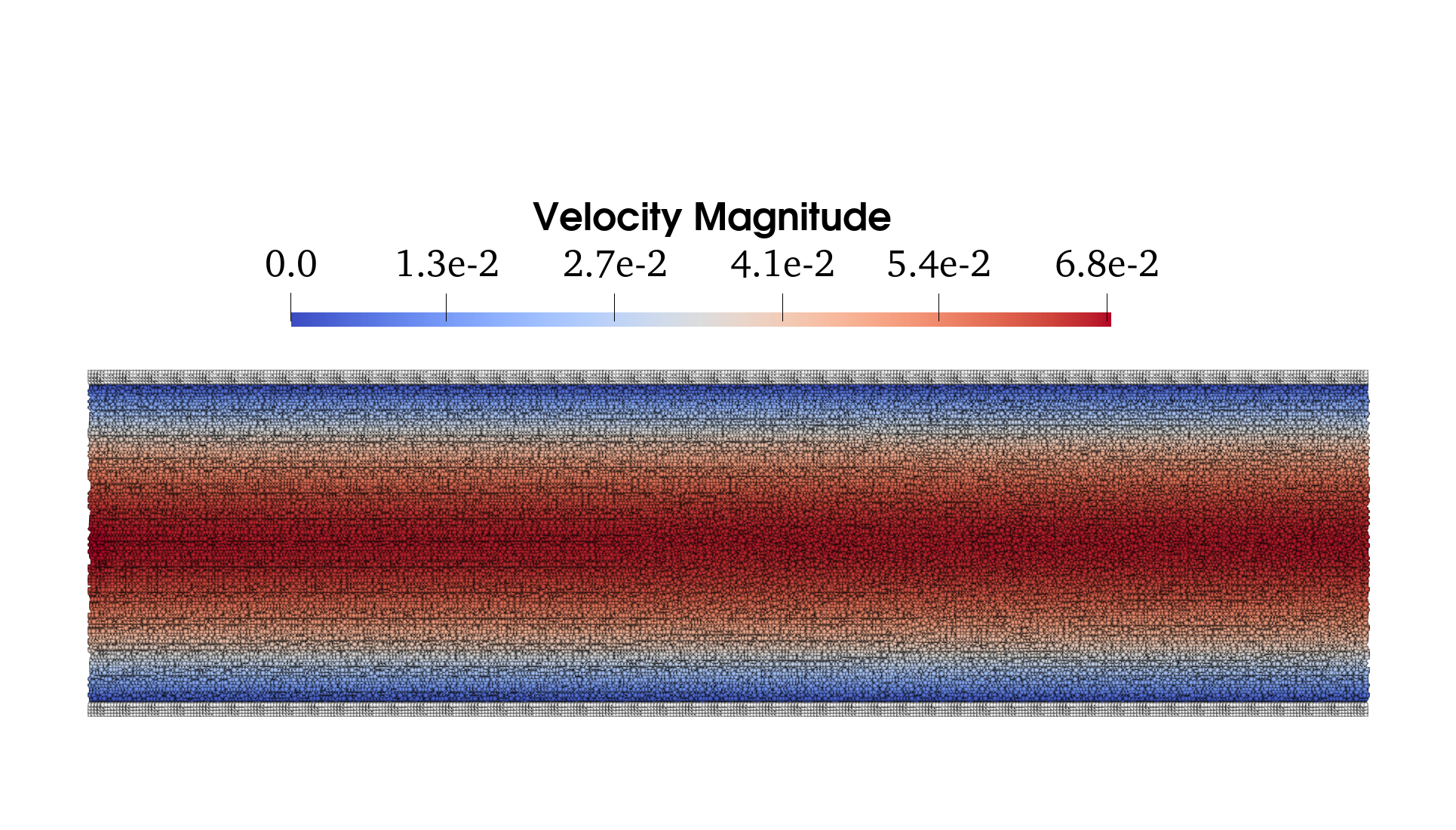}  
  \caption{\centering Velocity magnitude distribution at t5}
  \label{fig:7}
\end{figure}

\textcolor{blue}{\hyperref[fig:6]{Fig.~6}} and \textcolor{blue}{\hyperref[fig:7]{Fig.~7}} present the velocity–magnitude fields at the reference instants \(t_1\) and \(t_5\), characterizing the start-up phase and the steady state of the velocity field, respectively.


\subsection{Three-ring impact}
The three-ring impact benchmark is adapted from a prior self-contact study\cite{Yang2007}. Their reference setup features severe distortions of the medium ring and intermittent self contact.
In the present work we assess the 2D contact–impact performance on a three–ring configuration. The outermost ring is clamped along its exterior boundary. A small inner ring, initially located near a medium ring, is assigned a translational velocity 
$v = [30.0 \;\; 30.0]^T$ and impinges on the medium ring; after the two bodies travel together
toward the inside surface of the fixed outer ring. 
The geometry follows the reference setup: outer ring inner/outer diameters $(26,\,30)$, 
medium ring $(10,\,12)$, and small ring $(8,\,10)$, with centers at 
$[0,\,0]^T$, $[-7.9,\,8.5]^T$, and $[7.9,\,-8.5]^T$, respectively.
All rings are modeled as incompressible Neo-Hookean solids with Young’s module 
$E_1=10{,}000$ (small), $E_2=2{,}250$ (medium), and $E_3=288{,}000$ (outer),
a common Poisson ratio $\nu=0.125$, and reference densities 
$\rho_1=0.1$, $\rho_2=0.01$, and $\rho_3=1.0$.  
The simulation exhibits large elastic deformation and intermittent self-contact in the medium ring. 
At the output times reported in the reference\cite{Yang2007}, our SPHBody-based simulation reproduces the same qualitative features as the benchmark.

\begin{figure}
  \centering
  \includegraphics[width=1.0\linewidth]{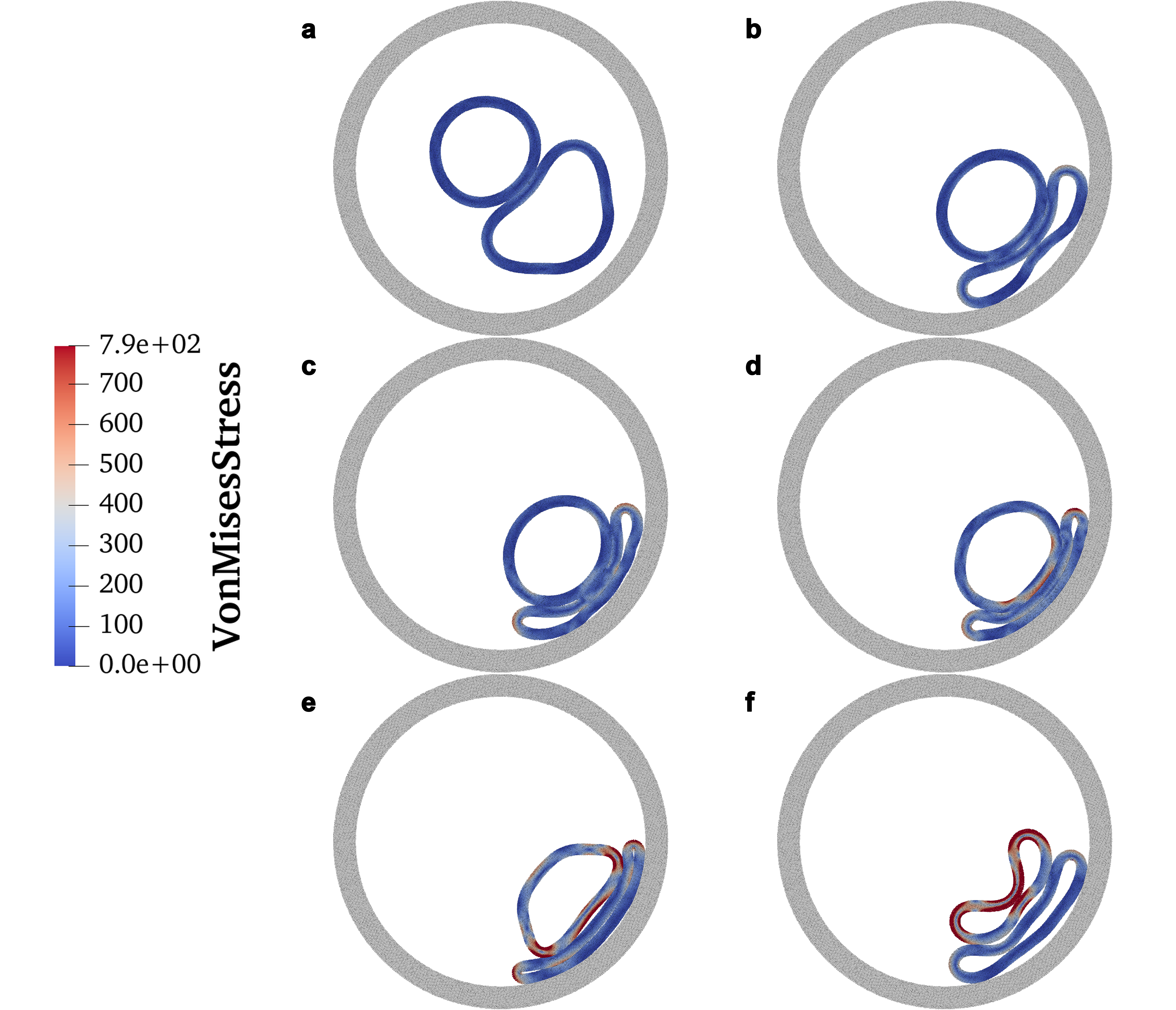}  
  \caption{Deformed configurations of the 2D impact problem at diﬀerent time steps using SPHBody}
  \label{fig:8}
\end{figure}

As shown in \textcolor{blue}{\hyperref[fig:8]{Fig.8}} (subfigures a–d), the deformation patterns and stress contours of the three interacting rings are in close agreement with the contact topology presented in the reference results.

In addition to presenting the deformation at each time step, we also provide the corresponding Von-Mises stress fields, thereby offering a more comprehensive view of the structural response. Furthermore, two additional time steps (e,f) are included to illustrate the subsequent interaction dynamics beyond those originally reported. 

\section{Coronary Artery Stent Implantation}
This section examines how stent implantation changes coronary hemodynamics. We first use a patient-specific coronary artery model with a lesion part to simulate the flow, and after fully expanding the stent, another simulation is conducted based on the modified arterial model. Finally, the effects on hemodynamics and vascular wall stress distribution are analyzed by comparing the simulation results before and after stent implantation. Fractional flow reserve (FFR) is computed to quantify functional improvement, and its post-stenting change is analyzed; combined maps of blood flow, pressure, and vessel-wall stress support clinical decisions.

\subsection{Parameter Setup}

Based on the patient's coronary geometries: The diameter of the left main coronary artery is $4$\,mm;  and the left anterior descending artery and left circumflex artery diameters ranges from $2$ to $4$\,mm. Vessel wall thickness ranges from $0.55$ to $1.0$\,mm and increases with branch diameter, larger branches have thicker walls. Blood rheology and vessel material constants are taken to be close to carotid-artery values \cite{villa2020coronary}. \textcolor{blue}{\hyperref[tab:coronary_parameters]{Table III}} lists the parameters used.

\begin{table*}[!t]
\caption{Parameters for the Coronary Artery Model, including vessel wall properties.}
\label{tab:coronary_parameters}
\centering
\begin{adjustbox}{angle=0,max width=0.75\textheight} 
\begin{ruledtabular}
\begin{tabular}{@{}p{0.28\textwidth}p{0.16\textwidth}p{0.28\textwidth}p{0.16\textwidth}@{}}
\textbf{Parameter} & \textbf{Value} & \textbf{Parameter} & \textbf{Value} \\
\hline
Diameter ($d_{\text{inlet}}$) & 4 mm & Density of vessel wall ($\rho_{\text{vessel}}$) & 1080 kg/m$^3$ \\
Vessel wall thickness ($\text{wall\_thickness}$) & 0.55--1 mm & Poisson's ratio ($\nu_{\text{vessel}}$) & 0.45 \\
Steady-state inlet flow velocity ($U_f$) & 0.315 m/s & Young's modulus ($E_{\text{vessel}}$) & $1.0\times10^{6}$ Pa \\
Density of blood ($\rho_{\text{blood}}$) & 1060 kg/m$^{3}$ & Bulk modulus ($K_{\text{vessel}}$) & $2.5\times10^{6}$ Pa \\
Dynamic viscosity ($\mu_{\text{blood}}$) & 0.0035 Pa$\cdot$s & Shear modulus ($G_{\text{vessel}}$) & $3.45\times10^{5}$ Pa \\
Reynolds number (Re) & 600 \\
\end{tabular}
\end{ruledtabular}
\end{adjustbox}
\end{table*}

\begin{figure}
    \centering
    \includegraphics[width=0.8\linewidth]{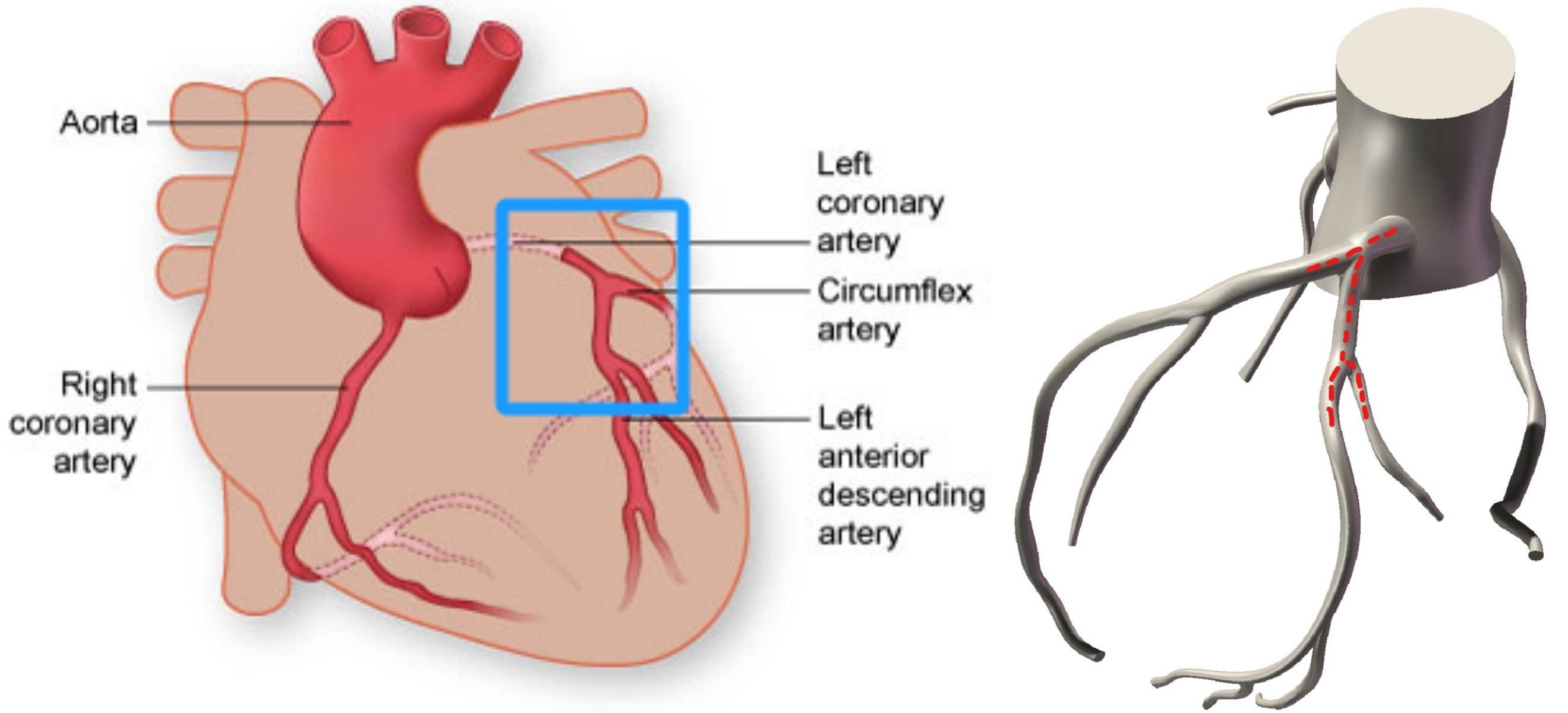}
    \caption{physiological location of the modeled object, The blue box represents the actual physiological position of the model (left), coronary artery model used as a reference for modeling (right).}
    \label{fig:9}
\end{figure}

\textcolor{blue}{\hyperref[fig:9]{Fig.~9}} (left) indicates the in vivo location of the modeled segment; the right panel shows the reference coronary geometry, with the target region highlighted by the red dashed box.

\begin{figure}
    \centering
    \includegraphics[width=0.8\linewidth]{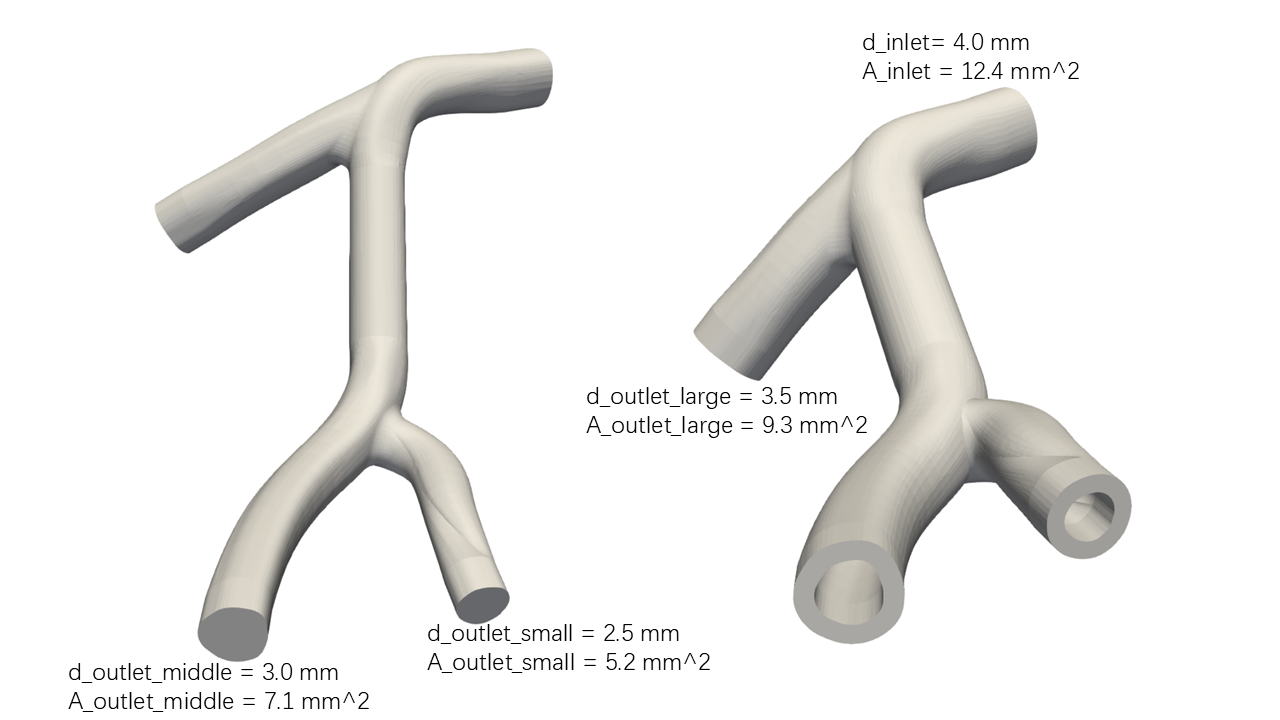}
    \caption{3D view of the STL-based coronary geometry for fluid (left) and vessel wall (right). One inlet and three outlets are labeled with diameters and areas.}
    \label{fig:10}
\end{figure}

\textcolor{blue}{\hyperref[fig:10]{Fig.~10}} presents a 3D schematic of the coronary model. The domain comprises one inlet and three outlets with distinct diameters: inlet $4.0$\,mm; outlets (large, medium, small) $3.5$\,mm, $3.0$\,mm, and $2.5$\,mm. The left panel is the fluid (blood) model; the right panel is the solid vessel-wall model.

\subsection{Simulation of Pre-Stent Coronary Artery}
We tested three spatial resolutions ($\Delta p=0.2,\,0.3,\,0.4$ mm) and observed negligible differences among the results shown in \textcolor{blue}{\hyperref[fig:12]{Fig.~12}} and \textcolor{blue}{\hyperref[fig:13]{Fig.~13}}. All three different resolutions show convergence at infinite time step. We report only the finest-resolution data here to avoid redundancy. Firstly, we defined the computational domain from the geometry model and set key simulation parameters, then imported geometries and obtained the distribution shown below. The steady inlet velocity is set to 0.315 m/s\cite{Kung2018FluentCases}. Velocity, pressure, and volumetric flow rate are recorded over time.

\begin{figure}
    \centering
    \includegraphics[width=1.0\linewidth]{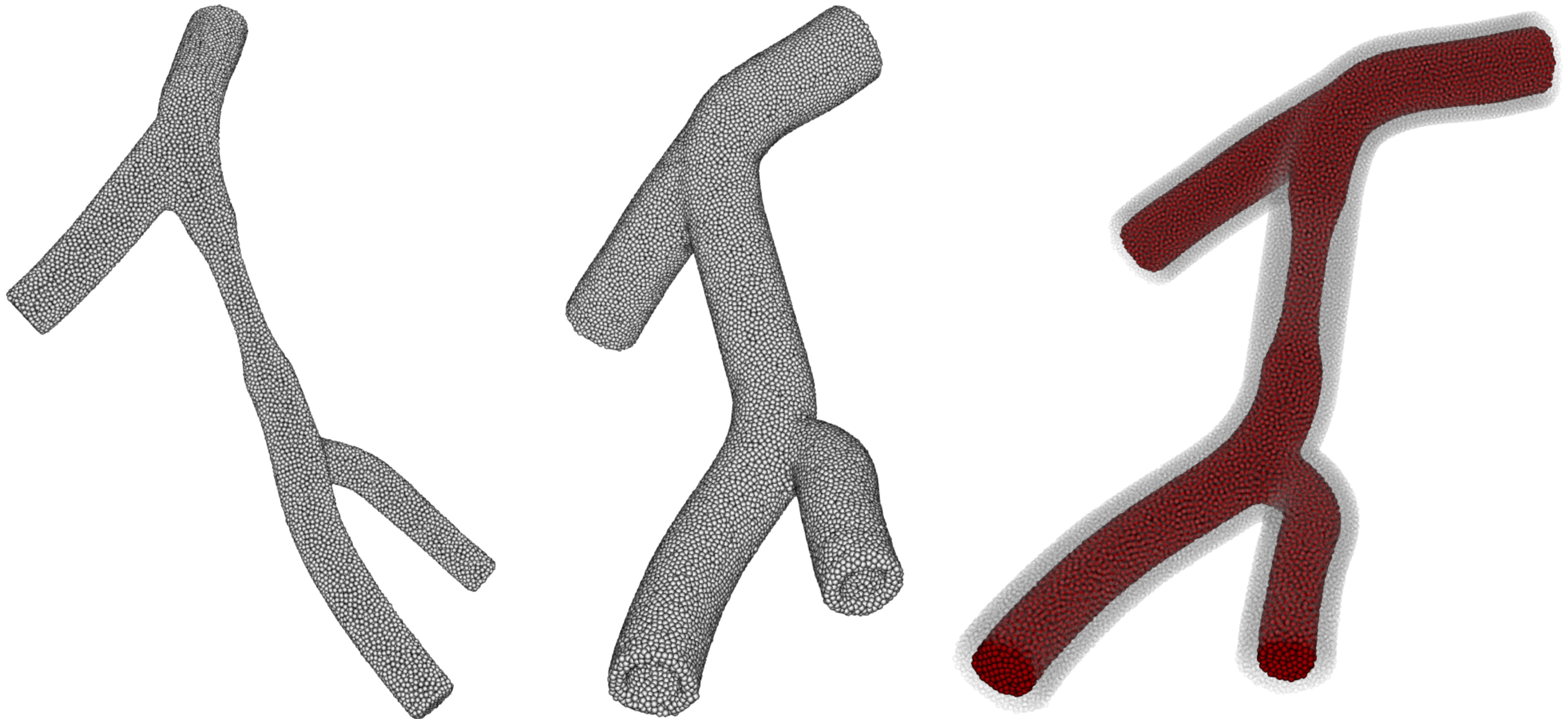}
    \caption{3D particle distribution of the coronary model. Left: blood (fluid). Middle: vessel wall (solid). Right: assembled model (transparent wall; red core indicates blood).}
    \label{fig:11}
\end{figure}

\begin{figure}
    \centering
    \includegraphics[width=1.0\linewidth]{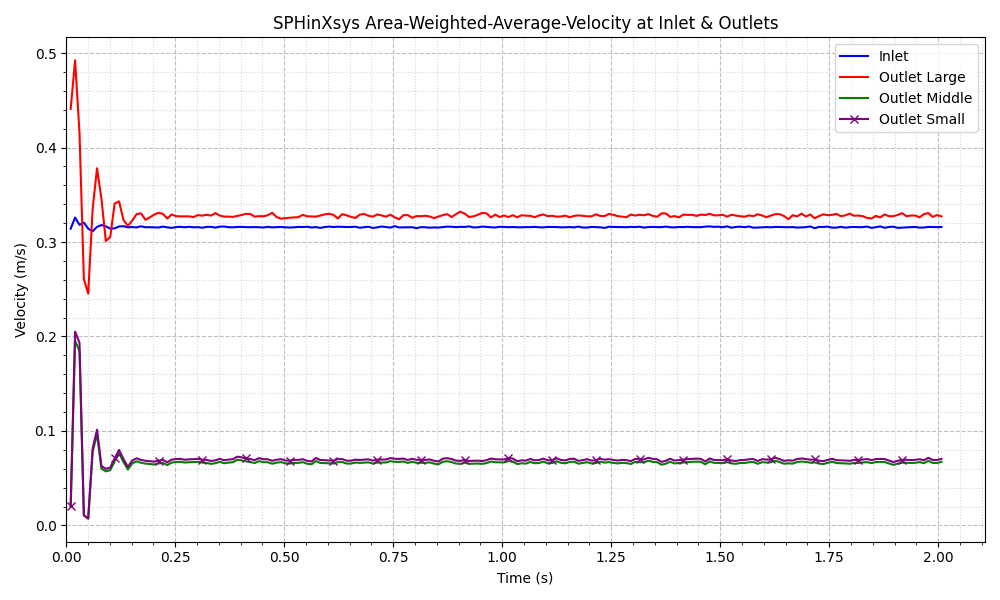}
    \caption{Area-weighted mean velocity versus time at the inlet and the three outlets. The inlet velocity (blue) and large outlet velocity (red) reach higher, stable values; the middle (green) and small (purple) outlets are lower and stabilize after initial fluctuations.}
    \label{fig:12}
\end{figure}

\textcolor{blue}{\hyperref[fig:12]{Fig.~12}} shows that the inlet area-weighted mean velocity quickly settles near $0.315$\,m/s and remains steady. The large outlet exhibits early oscillations, then stabilizes around $0.325$\,m/s, indicating it is the dominant outflow. The middle outlet shows mild transients and levels near $0.066$\,m/s, while the small outlet stabilizes near $0.07$\,m/s. Thus, most flow exits through the large branch, with weaker discharge through the other two, consistent with expected branch distribution.

\begin{figure}
    \centering
    \includegraphics[width=1.0\linewidth]{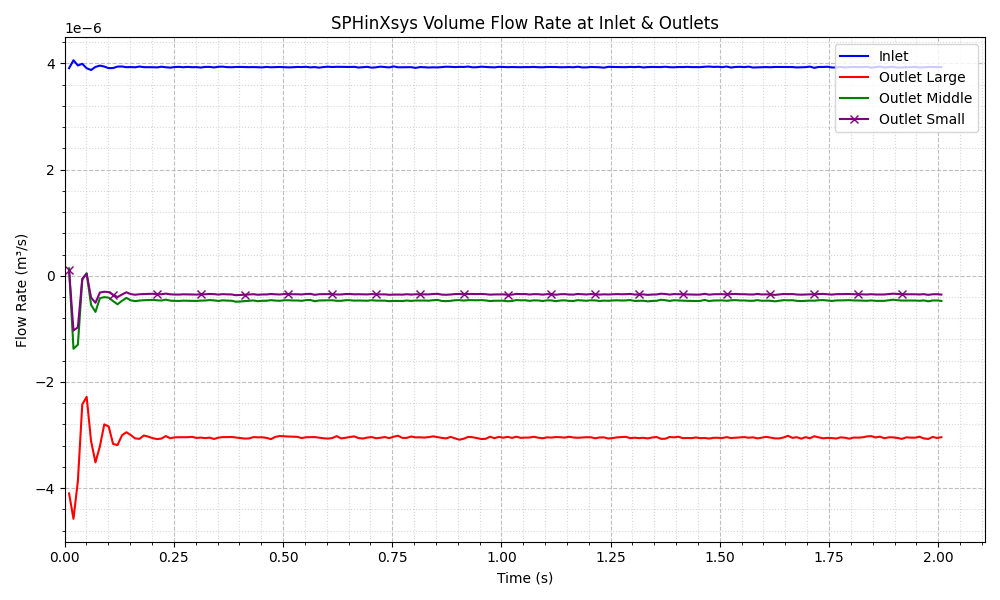}
    \caption{Volumetric flow rates at the inlet and three outlets. The inlet flow rate (blue) rapidly reaches a steady value. The large outlet flow rate (red) shows an initial overshoot and sign consistent with outflow; the middle (green) and small (purple) outlets carry smaller discharge.}
    \label{fig:13}
\end{figure}

\textcolor{blue}{\hyperref[fig:13]{Fig.~13}} shows that the inlet volumetric flow rate stabilizes near $3.9\times10^{-6}\,\text{m}^3/\text{s}$. The large outlet exhibits larger initial fluctuations before settling; negative values denote outward flow direction. The middle and small outlets display smaller amplitudes. Mass conservation holds: the inlet flow equals the sum of outlet flows. Overall, the large outlet conveys the majority of blood, and the other branches carry less after transients decay.

\begin{figure}
    \centering
    \includegraphics[width=1.0\linewidth]{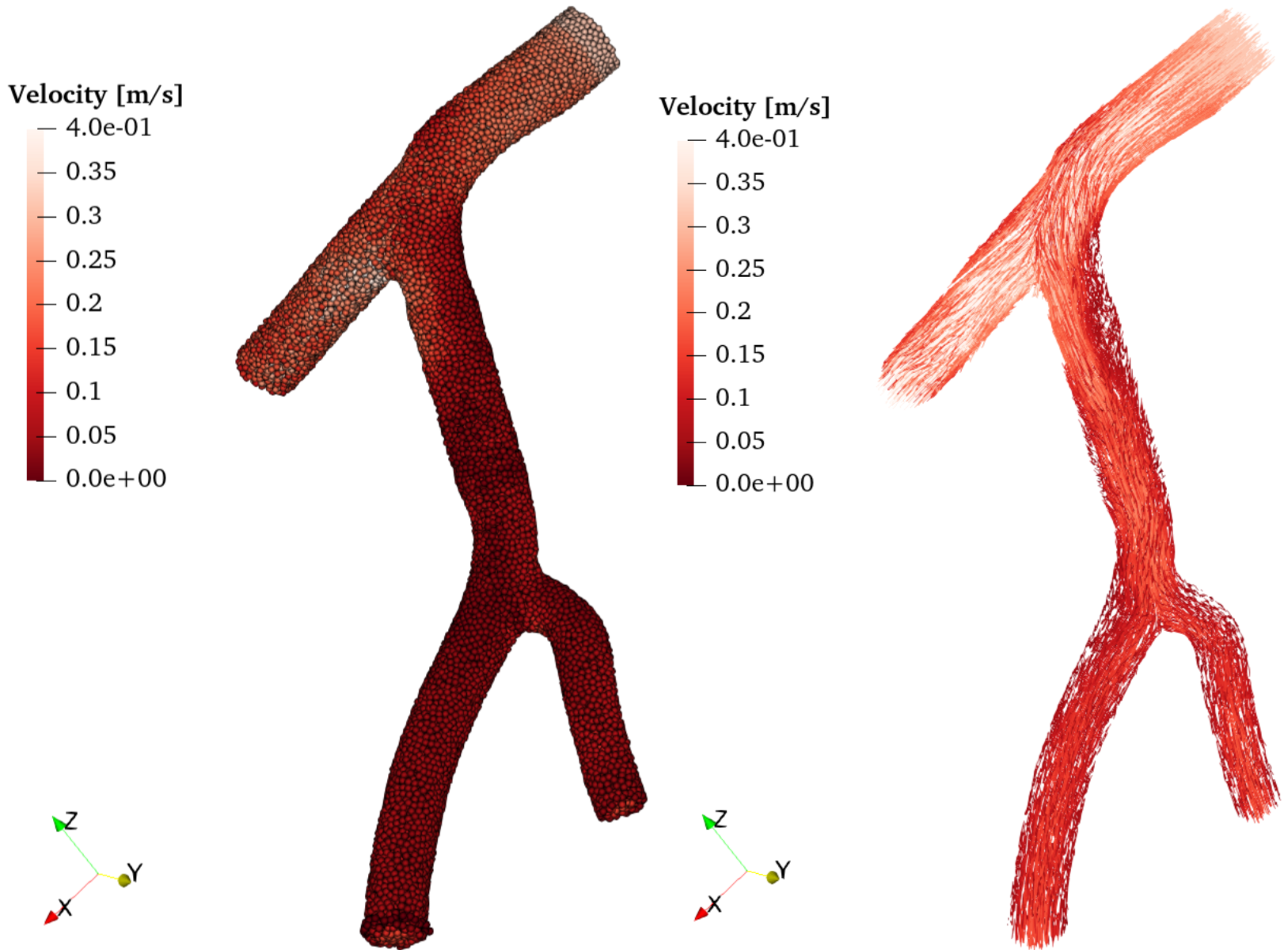}
    \caption{Left: velocity magnitude (bright = high). Right: streamlines indicating flow paths and local speeds.}
    \label{fig:14}
\end{figure}

\textcolor{blue}{\hyperref[fig:14]{Fig.~14}} visualizes the 3D velocity field. Higher speeds appear at the inlet and large outlet, while the middle and small outlets show lower velocities downstream of the stenosis. Flow accelerates through the narrowed segment, as expected from area reduction, and slows downstream as the lumen widens. Near-wall velocities are lower, reflecting boundary-layer effects.

The stress distribution on the coronary artery wall shows clear concentrations at the inlet and bifurcation. The fluid pressure is higher near the inlet and inner bifurcation walls and decreases toward the outlets. Correspondingly, Von-Mises stress on the wall peaks near the bifurcation and inlet, where flow loading and geometric curvature are stronger. Regions with lower velocities (distal branches, away from the bifurcation) exhibit reduced wall stress.

\begin{figure}
    \centering
    \includegraphics[width=1.0\linewidth]{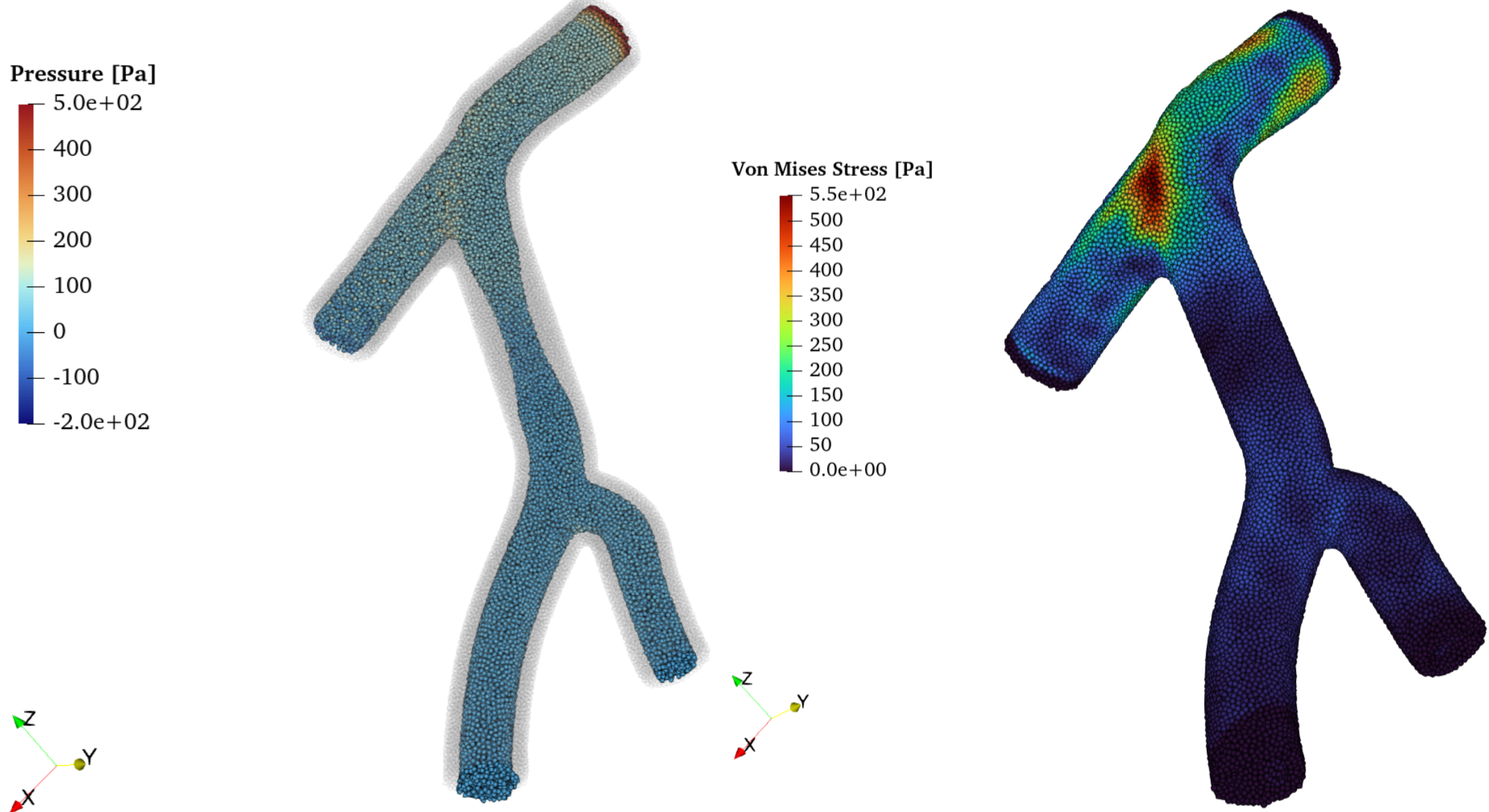}
    \caption{Left: pressure distribution in the fluid (warm colors = high). Right: Von-Mises stress in the wall (red = high, blue = low).}
    \label{fig:15}
\end{figure}

\begin{figure}
    \centering
    \includegraphics[width=1.0\linewidth]{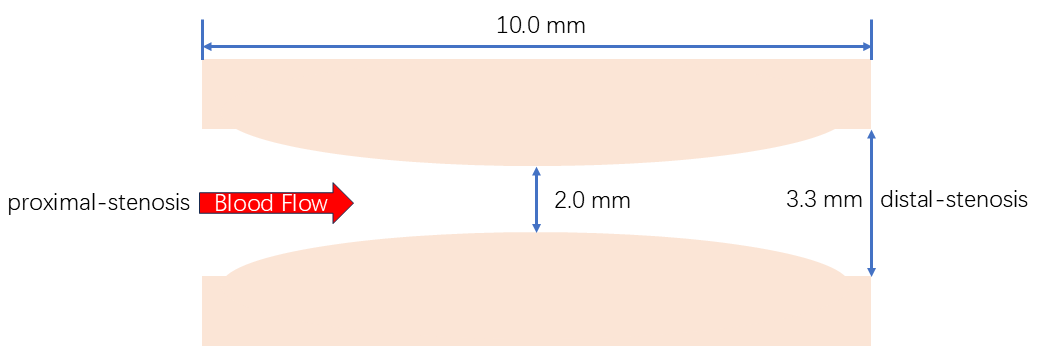}
    \caption{Geometry of the stenosis showing proximal, stenotic, and distal segments. The lumen narrows from $3.3$\,mm (non-stenotic) to $2.0$\,mm across a $10$\,mm length.}
    \label{fig:16}
\end{figure}

\begin{figure}
    \centering
    \includegraphics[width=1.0\linewidth]{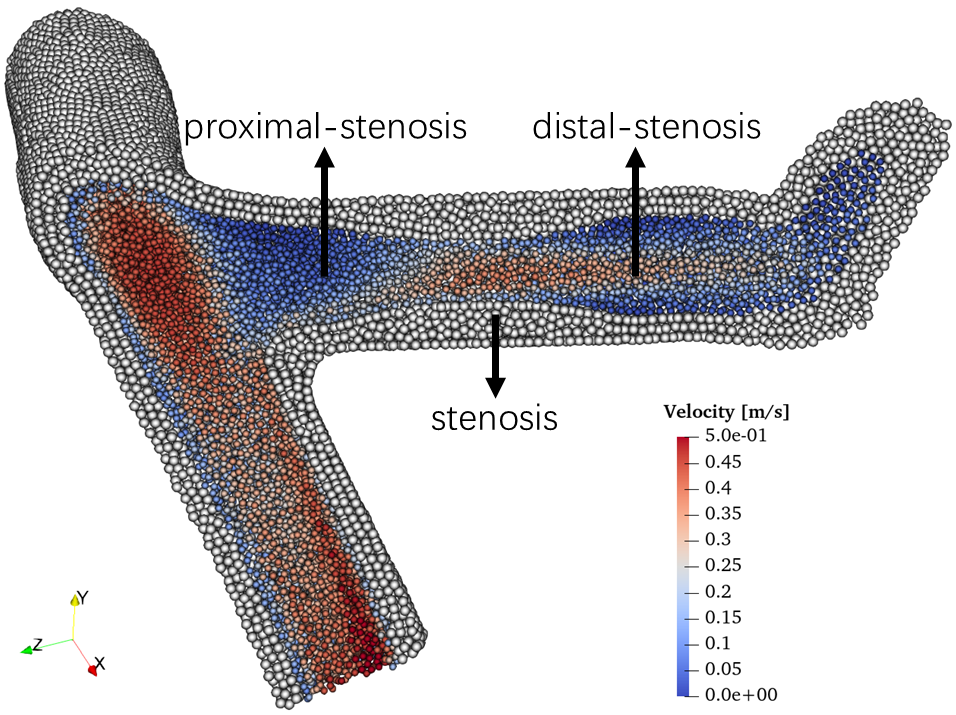}
    \caption{Cross-sectional velocity near the stenosis (red = high, blue = low). The arrows mark the locations before and after the stenosis.}
    \label{fig:17}
\end{figure}

The velocity map in \textcolor{blue}{\hyperref[fig:17]{Fig.~17}} shows moderate, more uniform speeds proximally; acceleration within the narrowed segment by continuity; and deceleration distally as the area recovers. Consistent with the global behavior, the main outlet carries higher velocity and flow than the middle and small branches.

\begin{figure}
    \centering
    \includegraphics[width=1.0\linewidth]{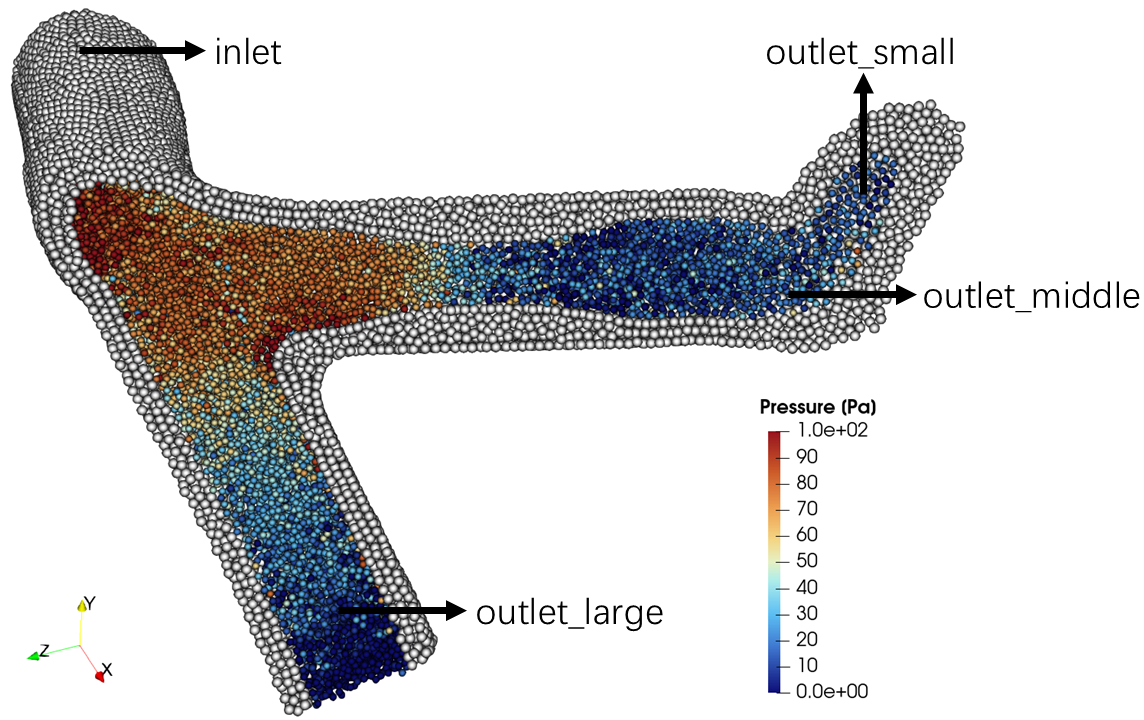}
    \caption{Cross-sectional pressure near the stenosis (red and yellow  = high, blue and lighter = low). The arrows indicate the flow direction, marking the inlet and different outlet regions.}
    \label{fig:18}
\end{figure}

\textcolor{blue}{\hyperref[fig:18]{Fig.~18}} depicts the pressure field: elevated near the inlet, a marked drop across the stenosis due to acceleration (Bernoulli effect), and partial recovery downstream. Among outlets, the large branch exhibits the lowest pressure, consistent with its greater flow, while the smaller branches show slightly higher pressures.

Fractional Flow Reserve (FFR) is an important indicator for evaluating whether coronary
artery stenosis leads to myocardial ischemia. It plays a crucial role, especially in assessing
the effectiveness of stent implantation before and after the procedure. FFR is the ratio of the
maximum blood flow pressure measured at the distal end (distal-stenosis) to the proximal
end (proximal-stenosis) of the stenotic lesion, reflecting the impact of vascular stenosis on
myocardial blood supply \cite{pijls1996measurement}. It is defined as:
\begin{equation}
\text{FFR} = \frac{P_d}{P_a},
\label{eq:ffr}
\end{equation}

\noindent\textbf{Where}
\begin{itemize}
    \item $P_d$: distal pressure under maximal hyperemia at end of the stenosis.
    \item $P_a$: proximal (usually aortic) pressure at end of the stenosis.
\end{itemize}

We computed distal-to-proximal pressure ratios before and after stenosis and found a stenotic-region FFR of about $0.45$. Interpretation is summarized in \textcolor{blue}{\hyperref[tab:ffr_assessment]{Table IV}}, values $<0.8$ indicate functionally significant disease, supporting stent implantation to restore perfusion and reduce ischemic risk.

\begin{table}
\centering
\caption{FFR Preoperative and Postoperative Assessment Criteria\cite{Agarwal2017FFRGrayZone}.}
\label{tab:ffr_assessment}
\begin{ruledtabular}
\begin{tabular}{lp{8cm}}
\textbf{FFR Assessment} & \textbf{Criteria} \\
\hline
\multicolumn{2}{c}{\textbf{Preoperative Evaluation}} \\\\
FFR $< 0.75$ & \textbf{Significant ischemia}, stent implantation \\ & required. \\\\
$0.75 \leq$ FFR $\leq 0.80$ & \textbf{Gray zone}, comprehensive judgment \\ & based on symptoms and imaging. \\\\
FFR $> 0.80$ & \textbf{No intervention needed} \\ & (unless symptoms are evident). \\\\
\multicolumn{2}{c}{\textbf{Postoperative Evaluation}} \\\\
Post-FFR $> 0.90$ & \textbf{Ideal stent implantation outcome,}\\ &sufficient blood flow recovery. \\\\
Post-FFR $\leq 0.90$ & \textbf{Possible residual stent issues}, including \\ &under-expansion, incomplete wall apposition, \\ & or uncovered lesions requiring optimization. \\
\end{tabular}
\end{ruledtabular}
\end{table}

\subsection{Simulation of Stent Implantation in Coronary Artery}
This subsection simulates the clinical stenting procedure (balloon angioplasty). Based on the surgical plan, the stent is positioned across the stenosis using prescribed translations and rotations extracted from the vessel wall model used in the prior flow study. Balloon inflation is modeled as a uniform pressure acting on the stent struts and is implemented equivalently as a radially outward load on the stent. Under this load, the stent expands, establishes contact with the constricted wall, and dilates the lesion until the target lumen diameter is achieved. \textcolor{blue}{\hyperref[tab:stent_properties]{Table V}} summarizes the stent's structural and material inputs.  

\begin{table}
\caption{Material and Structural Properties of the Coronary Stent.}
\label{tab:stent_properties}
\centering
\begin{ruledtabular}
\begin{tabular}{ll}
\textbf{Parameter} & \textbf{Value} \\
\hline
Stent material & Nitinol / Stainless Steel \\
Stent diameter ($d_{\text{stent}}$) & 2.0 mm \\
Stent length ($L_{\text{stent}}$) & 10 mm \\
Strut thickness ($t_{\text{strut}}$) & 80 -- 150 $\mu$m \\
Poisson's ratio ($\nu_{\text{stent}}$) & 0.3 \\
Young's modulus ($E_{\text{stent}}$) & $1.0 \times 10^{11}$ Pa \\
Shear modulus ($G_{\text{stent}}$) & $4.23 \times 10^{10}$ Pa \\
Bulk modulus ($K_{\text{stent}}$) & $1.67 \times 10^{11}$ Pa \\
\end{tabular}
\end{ruledtabular}
\end{table}

\begin{figure}
    \centering
    \includegraphics[width=1.0\linewidth]{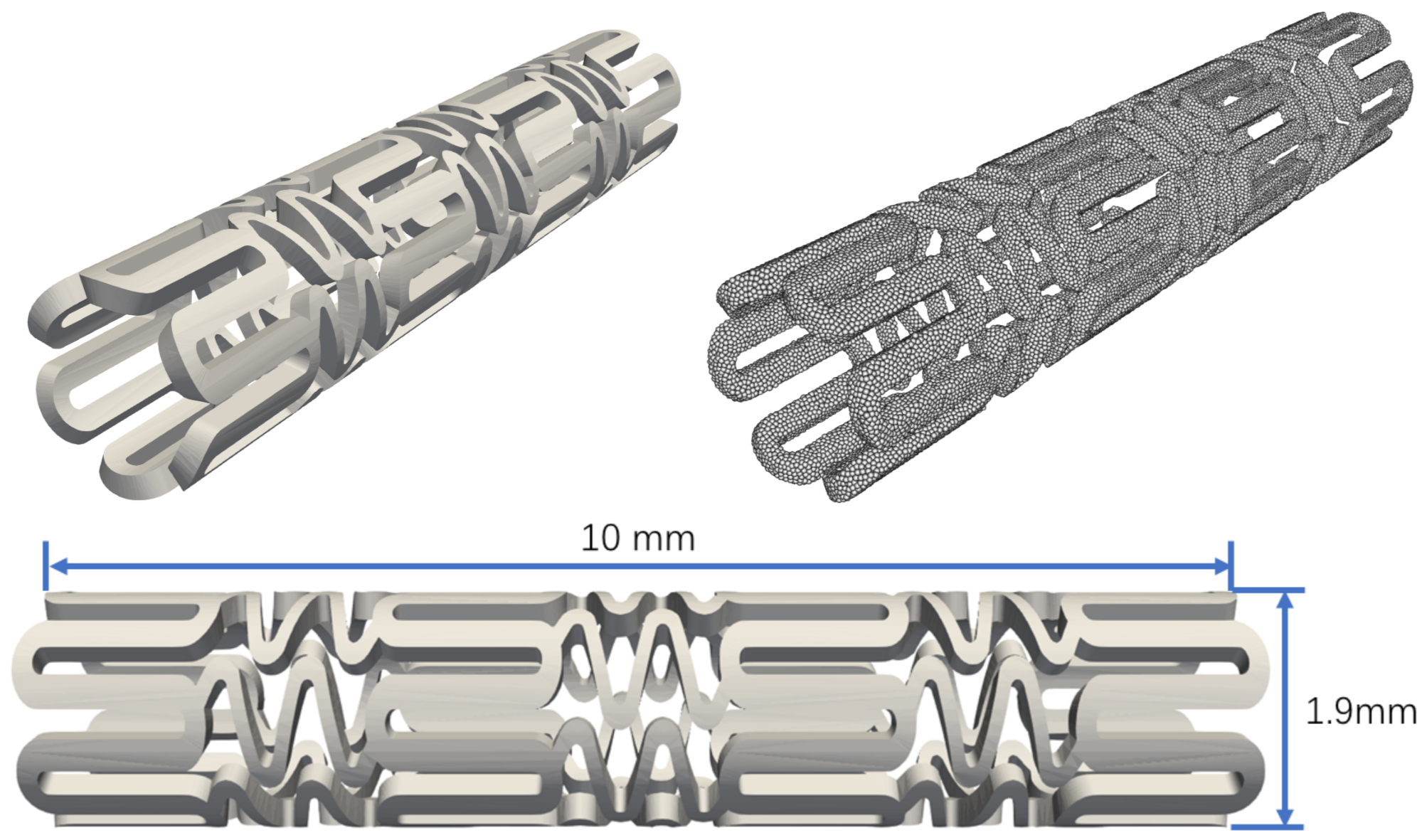}
    \caption{3D stent geometry (left), meshed/particle model (right), and side view with dimensions (bottom). The same stent structure is used for particle generation and subsequent simulation.}
    \label{fig:19}
\end{figure}

\textcolor{blue}{\hyperref[fig:19]{Fig.~19}} shows the imported geometry and corresponding particle discretization. The pattern is sized to the coronary lesion to yield a realistic expansion response. The initial stent diameter is 2\,mm; under radial forcing it dilates the lumen to about 3.3\,mm, i.e., the nominal vessel caliber. A real-time check of the stent’s bounding box regulates expansion, terminating the load once the prescribed diameter is reached. Because the stent is much finer than the global vessel scale, a multi-resolution scheme is adopted: the stent particle spacing is set to one-fifth of the vessel spacing to better capture strut deformation and wall contact.

\begin{figure}
    \centering
    \includegraphics[width=1.0\linewidth]{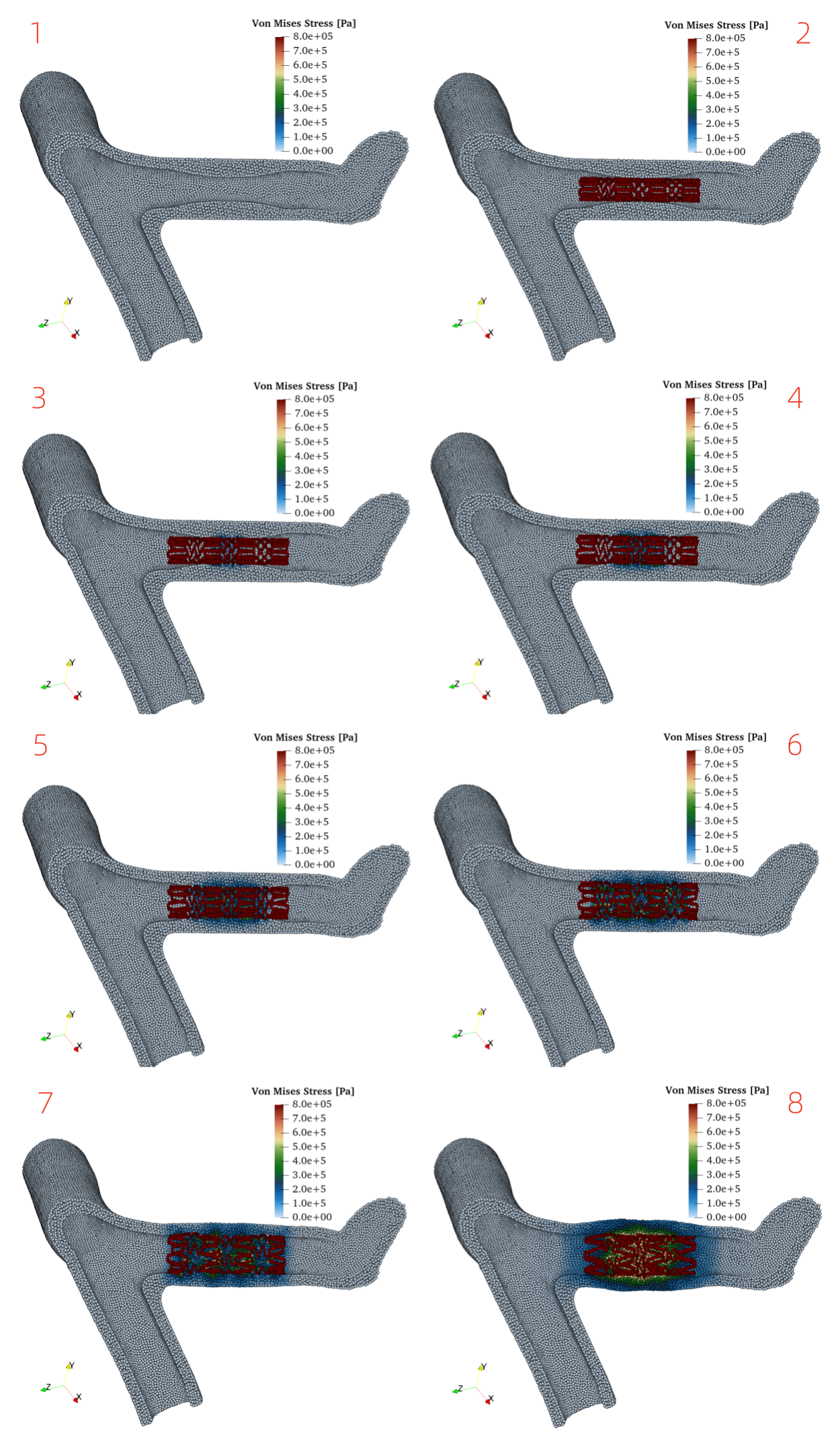}
    \caption{Stent deployment sequence (1–8): gradual expansion, vessel scaffolding, stenosis relief, and restoration of the target lumen diameter.}
    \label{fig:20}
\end{figure}

\textcolor{blue}{\hyperref[fig:20]{Fig.~20}} summarizes the deployment sequence. 
Panels are ordered left to right and top to bottom. 
The red structure denotes the stent, and the vessel color map encodes Von-Mises stress (darker = higher). 
Balloon angioplasty is simulated as a uniform pressure on the stent (implemented as a radially outward load), so darker regions indicate where the wall increasingly bears load as the stent expands. 
The sequence proceeds as follows:
\begin{itemize}
    \item \textbf{Initial state (Step 1):} Pre-implantation vessel in its original condition, with near-zero wall stress and nearly uniform coloration. The vessel geometry is unchanged.
    \item \textbf{Stent positioning (Step 2):} The stent is aligned across the stenotic lesion but remains unexpanded.
    \item \textbf{Gradual expansion (Steps 3--6):} The applied load increases the stent diameter, enlarging stent--wall contact. Wall stress redistributes with localized peaks where struts engage the wall; the stent begins to widen the lumen.
    \item \textbf{Full deployment (Steps 7--8):} The stent reaches its target diameter and conforms to the vessel wall. Stress in the stented segment peaks (darker colors denote higher Von-Mises stress), and the lumen is restored to the planned size.
\end{itemize}

Under radial loading, the stent is gradually expanded and the supported vessel wall reaches about $10^{6}\ \text{Pa}$, confirming the effectiveness of the stent in stabilizing the vessel\cite{Liu2022TaperedStentExpansion}.

\subsection{Simulation of Post-Stent Coronary Artery}
This subsection extends the previous analysis by using the vascular geometry after stent implantation and expansion at the stenotic region for further fluid dynamics simulations. To ensure strict comparability, we keep the numerical method, fluid properties, boundary conditions, and solver settings unchanged. This allows a direct assessment of how stenting modifies the hemodynamics relative to the pre-stent case.
\begin{table*}[!t]
\centering
\caption{Comparison of Flow and Velocity Before and After Stent Deployment.}
\label{tab:6}

\rotatebox{0}{%
  \resizebox{0.75\textheight}{!}{
  \begin{ruledtabular}
  \begin{tabular}{ll}
  \textbf{Parameter} & \textbf{Value Before $\to$ After} \\
  \hline
  Inlet Velocity ($v_{\mathrm{inlet}}$) [m/s] & 0.315 $\to$ 0.315 \\
  Large Outlet Velocity ($v_{\mathrm{outlet\_large}}$) [m/s] & 0.325 $\to$ 0.310 \\
  Middle Outlet Velocity ($v_{\mathrm{outlet\_middle}}$) [m/s] & 0.066 $\to$ 0.095 \\
  Small Outlet Velocity ($v_{\mathrm{outlet\_small}}$) [m/s] & 0.069 $\to$ 0.098 \\
  Pre-Stenosis Velocity ($v_{\mathrm{pre\_stenosis}}$) [m/s] & 0.108 $\to$ 0.115 \\
  Post-Stenosis Velocity ($v_{\mathrm{post\_stenosis}}$) [m/s] & 0.110 $\to$ 0.122 \\
  Inlet Flow Rate ($Q_{\mathrm{inlet}}$) [m$^{3}$/s] & $3.92\times10^{-6}$ $\to$ $3.93\times10^{-6}$ \\
  Large Outlet Flow Rate ($Q_{\mathrm{outlet\_large}}$) [m$^{3}$/s] & $-3.06\times10^{-6}$ $\to$ $-2.76\times10^{-6}$ \\
  Middle Outlet Flow Rate ($Q_{\mathrm{outlet\_middle}}$) [m$^{3}$/s] & $-4.75\times10^{-7}$ $\to$ $-6.75\times10^{-7}$ \\
  Small Outlet Flow Rate ($Q_{\mathrm{outlet\_small}}$) [m$^{3}$/s] & $-3.6\times10^{-7}$ $\to$ $-5.18\times10^{-7}$ \\
  Stenosis Flow Rate ($Q_{\mathrm{stenosis}}$) [m$^{3}$/s] & $8.5\times10^{-7}$ $\to$ $11.5\times10^{-7}$ \\
  Fractional Flow Reserve (FFR) & 0.45 $\to$ 0.91 \\
  \end{tabular}
  \end{ruledtabular}
}}
\end{table*}

\textcolor{blue}{\hyperref[tab:6]{Table VI}} summarizes the changes of velocity and flow redistribution due to stenting. The large outlet slows (0.325\,$\to$\,0.310\,m/s) and carries less flow ($-3.06\times10^{-6}\,\to\,-2.76\times10^{-6}$\,m$^3$/s), while the middle and small outlets gain velocity (0.066\,$\to$\,0.095\,m/s and 0.069\,$\to$\,0.098\,m/s) and flow, indicating a more balanced branch distribution after restoring the artery’s inner channel.

Local speeds near the lesion also rise modestly: $v_{\mathrm{pre\_stenosis}}$ increases from 0.108 to 0.115\,m/s and $v_{\mathrm{post\_stenosis}}$ from 0.110 to 0.122\,m/s, reflecting reduced resistance across the previously narrowed segment. The stenosis-region flow rate grows from $8.5\times10^{-7}$ to $1.15\times10^{-6}$\,m$^3$/s, confirming improved patency. Functionally, FFR improves from 0.45 to 0.91, consistent with restoration of adequate perfusion.

In summary, this table demonstrates the impact of stent deployment on hemodynamics. Although the inlet flow rate and velocity remain unchanged, the stent effectively reduces the
flow resistance in the stenotic region, leading to a more balanced blood flow distribution.
This increases the flow rate through the middle and small outlets while improving the blood
passage through the narrowed segment. These findings indicate that the stent is crucial in
restoring vascular patency.

\begin{figure}
    \centering
    \includegraphics[width=1.0\linewidth]{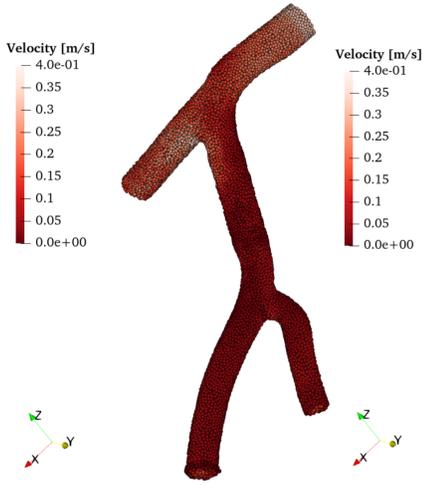}
    \caption{Post-stent velocity field (left: magnitude, bright = high; right: streamlines). Flow remains fast in the main trunk; near walls velocities are lower, and branch regions show redistribution consistent with a relieved stenosis.}
    \label{fig:21}
\end{figure}

\textcolor{blue}{\hyperref[fig:21]{Fig.~21}}  shows the post-stent velocity magnitude and streamlines. The main vessel maintains higher speeds (up to \(\sim\)0.4\,m/s) in straight segments; velocities reduce toward bifurcations as flow partitions among branches. Near-wall speeds are low, as expected from boundary-layer behavior. Streamlines are parallel in the trunk (stable flow) and curve through the bifurcation, where local shear and possible weak vortical structures appear, indicating redistribution among outlets.

\begin{figure}
    \centering
    \includegraphics[width=1.0\linewidth]{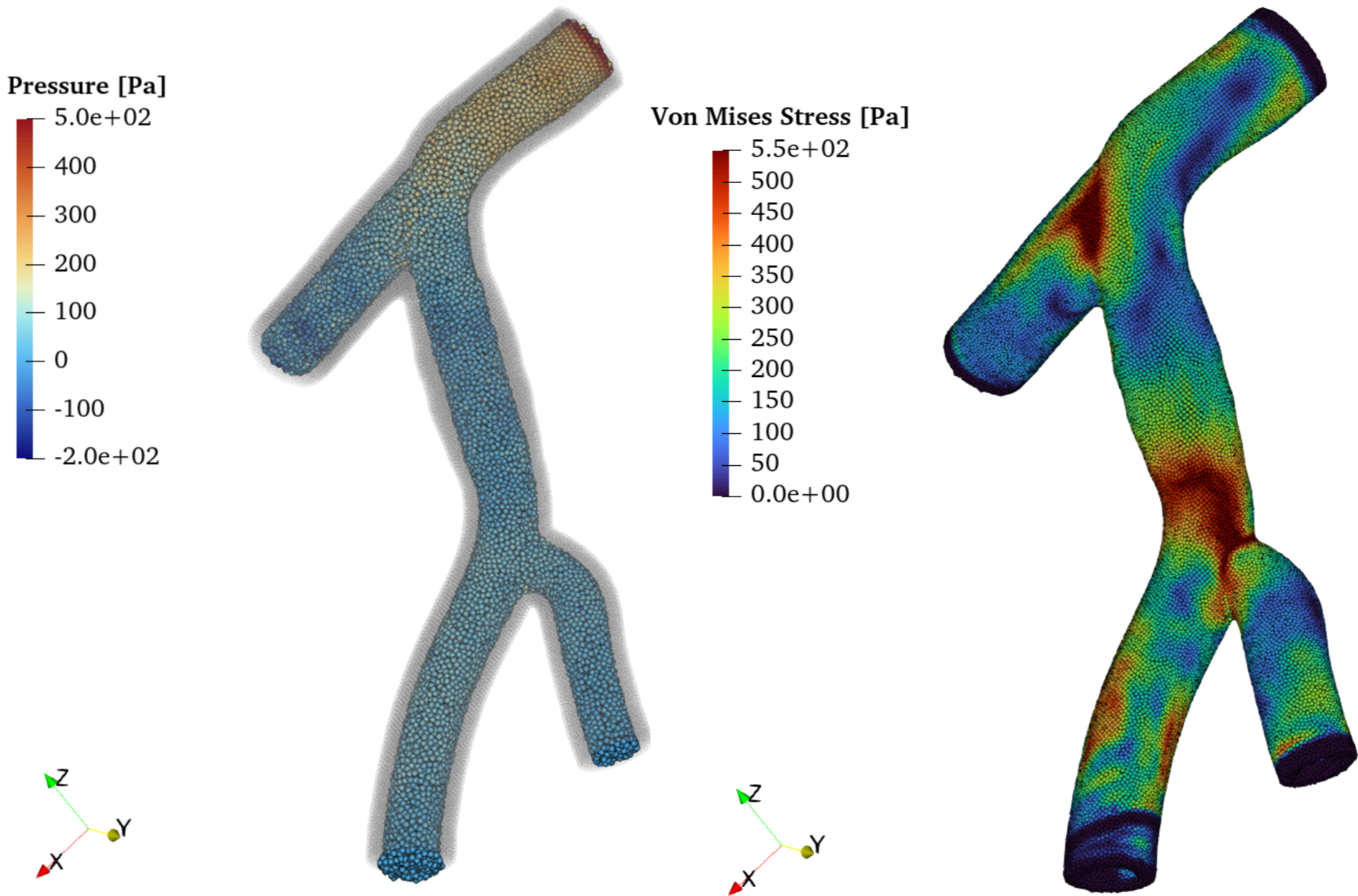}
    \caption{Post-stent hemodynamics: fluid pressure (left) and Von-Mises stress in the wall  (right)  (red and yellow  = high, blue and lighter = low). Elevated pressures appear proximally and fall distally; wall stress concentrates at curved segments and bifurcations.}
    \label{fig:22}
\end{figure}

\textcolor{blue}{\hyperref[fig:22]{Fig.~22}} presents the post-stent pressure field and the wall’s Von-Mises stress. Pressure is higher upstream in the main vessel and progressively falls downstream and across bifurcations, consistent with hemodynamic losses at junctions and former narrow sites. The right panel shows the wall stress pattern: stresses concentrate along curved segments and at bifurcations, where flow impingement, secondary motion, and shear are stronger; elsewhere the stress is lower and more uniform.
\begin{figure}
    \centering
    \includegraphics[width=1.0\linewidth]{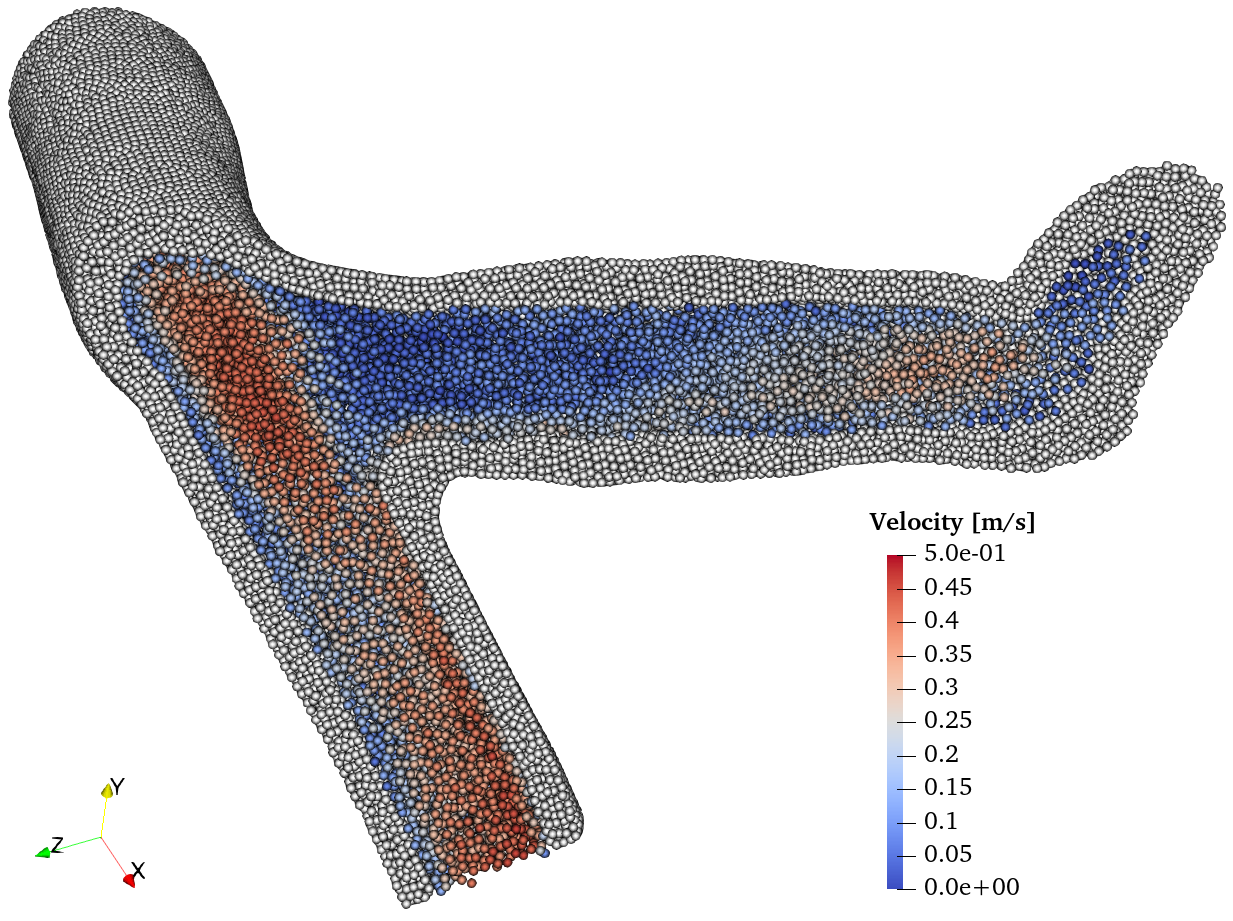}
    \caption{Cross-sectional velocity in the stenotic segment after stenting (red = high, blue = low). The restored inner channel supports smooth transit through the former constriction.}
    \label{fig:23}
\end{figure}

\textcolor{blue}{\hyperref[fig:23]{Fig.~23}} indicates that blood speed is highest in the main channel, particularly near the inlet, showing substantial kinetic energy at entry. Speeds decline through the bifurcation and farther downstream, visualized as blue and light-blue regions. This decrease reflects flow being redistributed among multiple outlets. In the treated (post-stent) narrowing, flow passes smoothly with little tendency to stagnate, which lowers the local pressure load at the former lesion and enhances regional perfusion.

\begin{figure}
    \centering
    \includegraphics[width=1.0\linewidth]{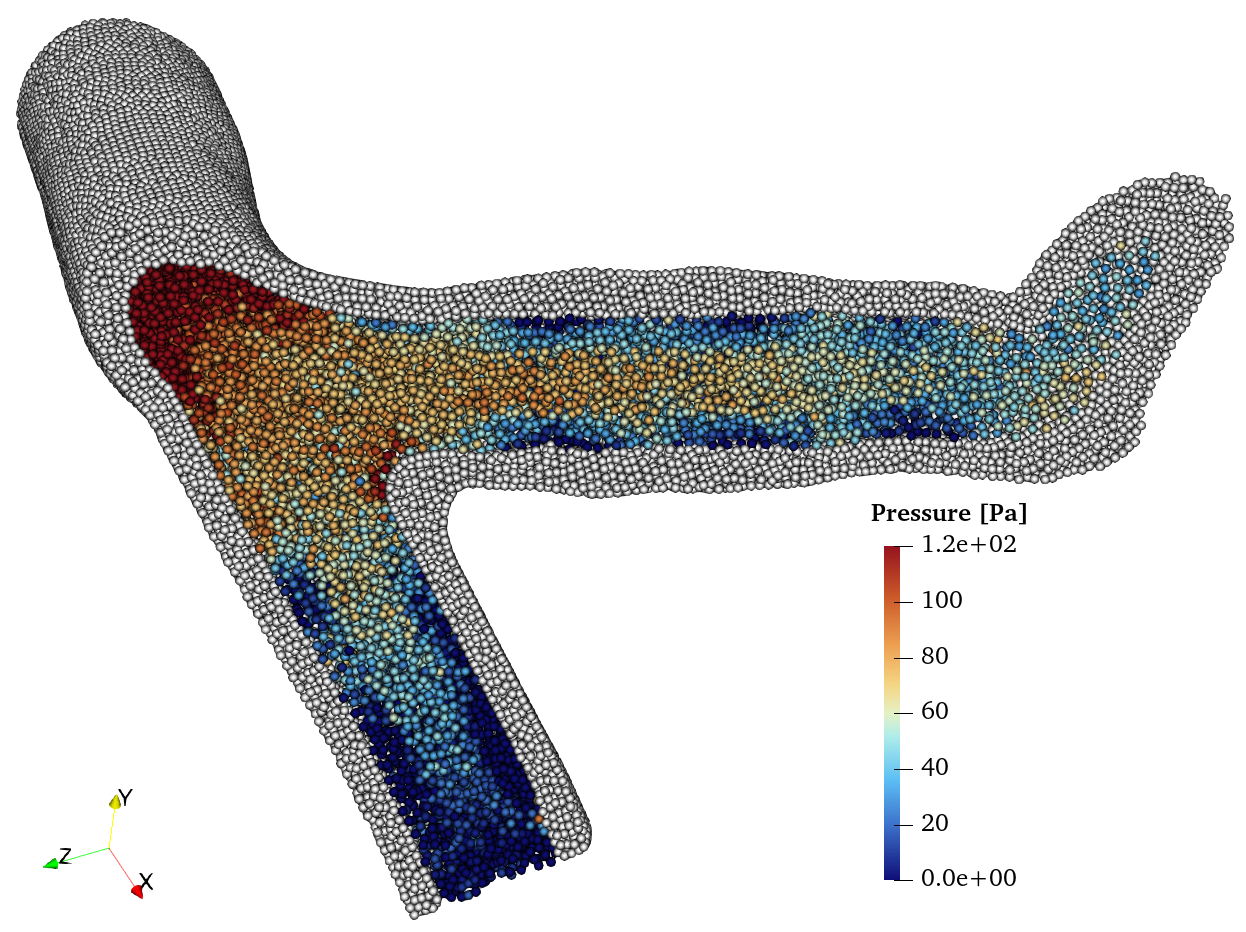}
    \caption{Cross-sectional pressure in the stenotic segment after stenting. Warm colors denote higher pressure; cooler colors mark lower pressure. Pressure is more uniform downstream, with no marked drop across the treated site.}
    \label{fig:24}
\end{figure}

\textcolor{blue}{\hyperref[fig:24]{Fig.~24}} shows the post-stent pressure field. Pressure peaks near the inlet and declines along the main vessel. A clear gradient forms at the bifurcation—higher at the junction, lower in distal branches—reflecting flow redistribution and local energy losses. In the previously narrowed segment, pressure remains stable without a marked drop, indicating improved patency and reduced loss. Downstream, pressures are more uniform with no evident buildup.

\begin{figure*}
    \centering
    \begin{minipage}{0.48\textwidth}
        \includegraphics[width=\linewidth]{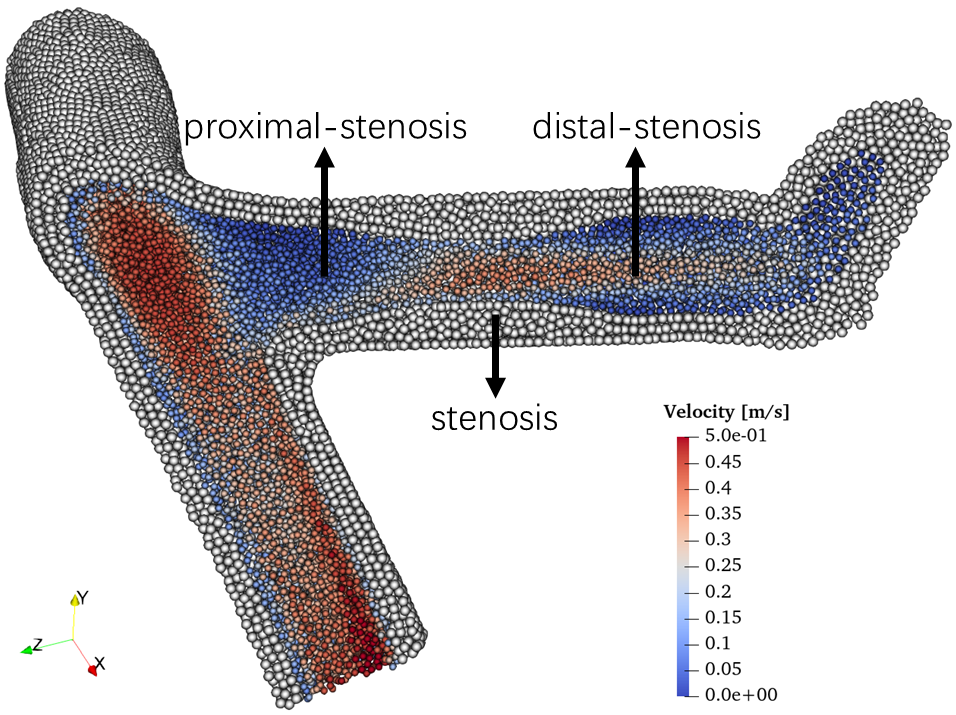}
        \vspace{4pt}
        \centering (a)
        \label{fig:front}
    \end{minipage}\hfill
    \begin{minipage}{0.48\textwidth}
        \includegraphics[width=\linewidth]{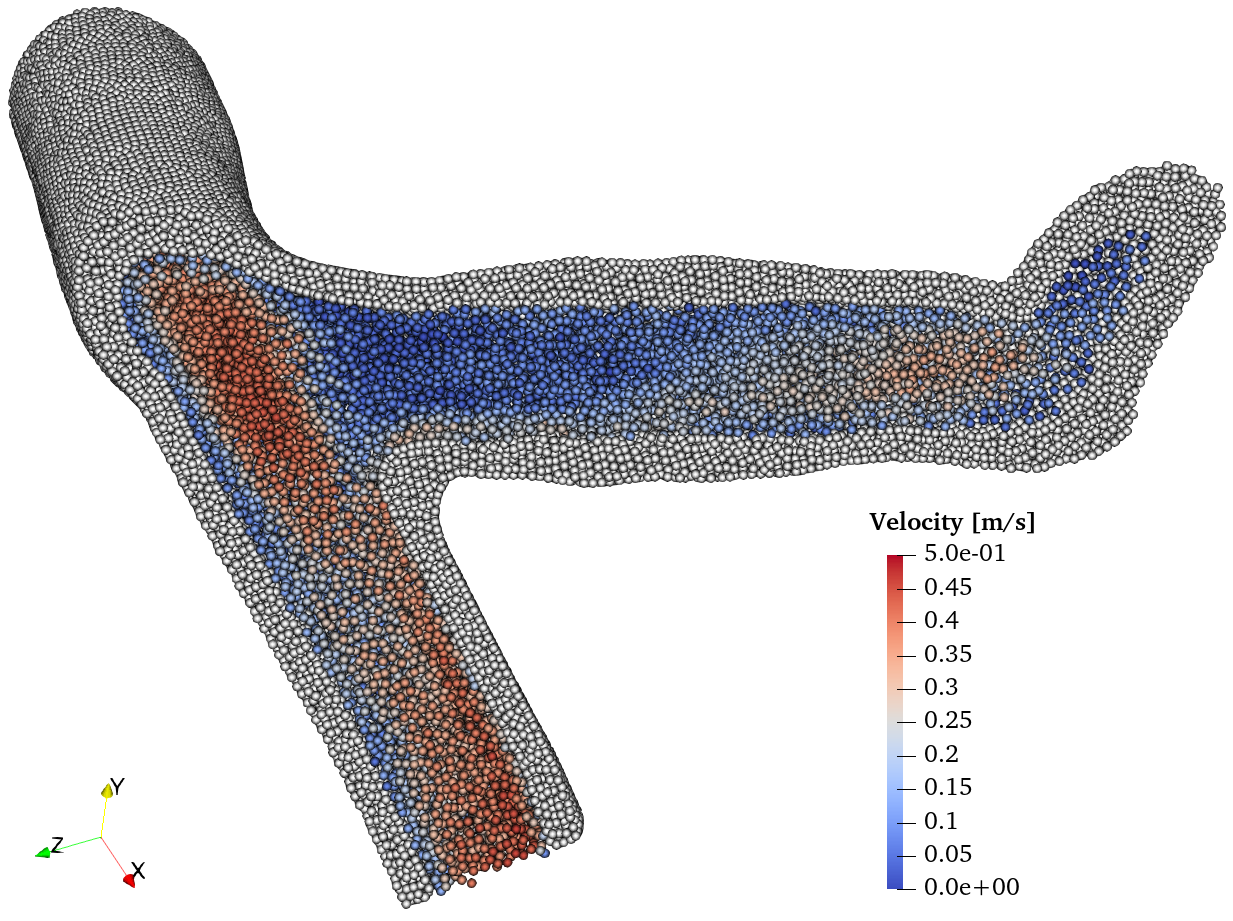}
        \vspace{4pt}
        \centering (b)
        \label{fig:3d}
    \end{minipage}
    \caption{Cross-sectional velocity in the stenotic region before (left) and after (right) stenting. Red indicates high speed; blue and light tones indicate low speed. Arrows mark proximal, stenosis, and distal segments.}
    \label{fig:25}
\end{figure*}

\textcolor{blue}{\hyperref[fig:25]{Fig.~25}}  contrasts pre-stent and post-stent velocity fields. Before deployment, the narrowed artery forces local acceleration and uneven profiles. After deployment, the velocity distribution becomes smoother across the section, with reduced extremes and improved through flow, matching the outlet redistribution and the increase in FFR.

\subsection{Results Analysis}

\subsection*{Changes in Blood Flow Velocity in the Stenotic Region}

\textbf{Before Stent Deployment (\textcolor{blue}{\hyperref[fig:25]{Fig.~25}} (a)):} The stenotic region (labeled as stenosis)
exhibits significant velocity changes. The velocity before the stenosis (proximal-stenosis) is relatively low (blue). However, as blood enters the stenotic region, the velocity sharply increases, forming a distinct high-speed channel (red). Subsequently, in the post-stenosis region, the velocity drops significantly (transitioning from red to blue).

\textbf{After Stent Deployment (\textcolor{blue}{\hyperref[fig:25]{Fig.~25}} (b)):} The velocity field becomes smoother and more uniform. The high-speed core within the lesion disappears, and overall variations are milder. This indicates reduced local  flow resistance and easier passage of blood through the treated segment.

\subsection*{Velocity Distribution Proximal and Distal to the Stenotic Region}

\textbf{Before Stent Implantation:}The velocity in the proximal-stenosis region is relatively low. In contrast, the distal-stenosis region experiences a significant velocity drop, indicating substantial pressure loss and flow velocity fluctuations caused by the stenosis.

\textbf{After Stent Implantation:}  The velocity transition across the stenosis becomes smoother after stent implantation, with no abrupt accelerations or decelerations. showing that resistance has been lowered and local circulation improved effectively.

\subsection*{Improvement in Flow Uniformity}

\textbf{Before Stent Implantation:} A tightly concentrated high-speed flow forms in the narrowed zone, reflecting strong local acceleration and potentially unstable patterns.

\textbf{After Stent Implantation:} The velocity gradient significantly diminishes, implying smoother passage, less turbulence, and a more regular shear-stress field.

\begin{figure*}
    \centering
    \begin{minipage}{0.48\textwidth}
        \includegraphics[width=\linewidth]{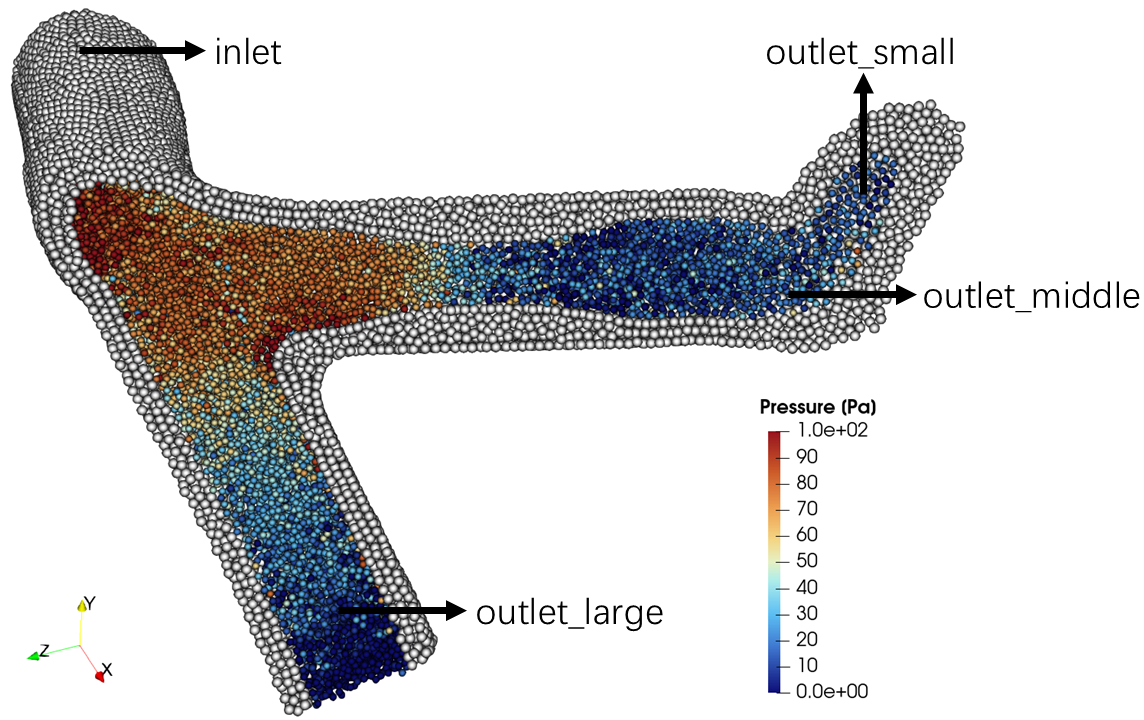}
        \vspace{4pt}
        \centering (a)
        \label{fig:26}
    \end{minipage}\hfill
    \begin{minipage}{0.48\textwidth}
        \includegraphics[width=\linewidth]{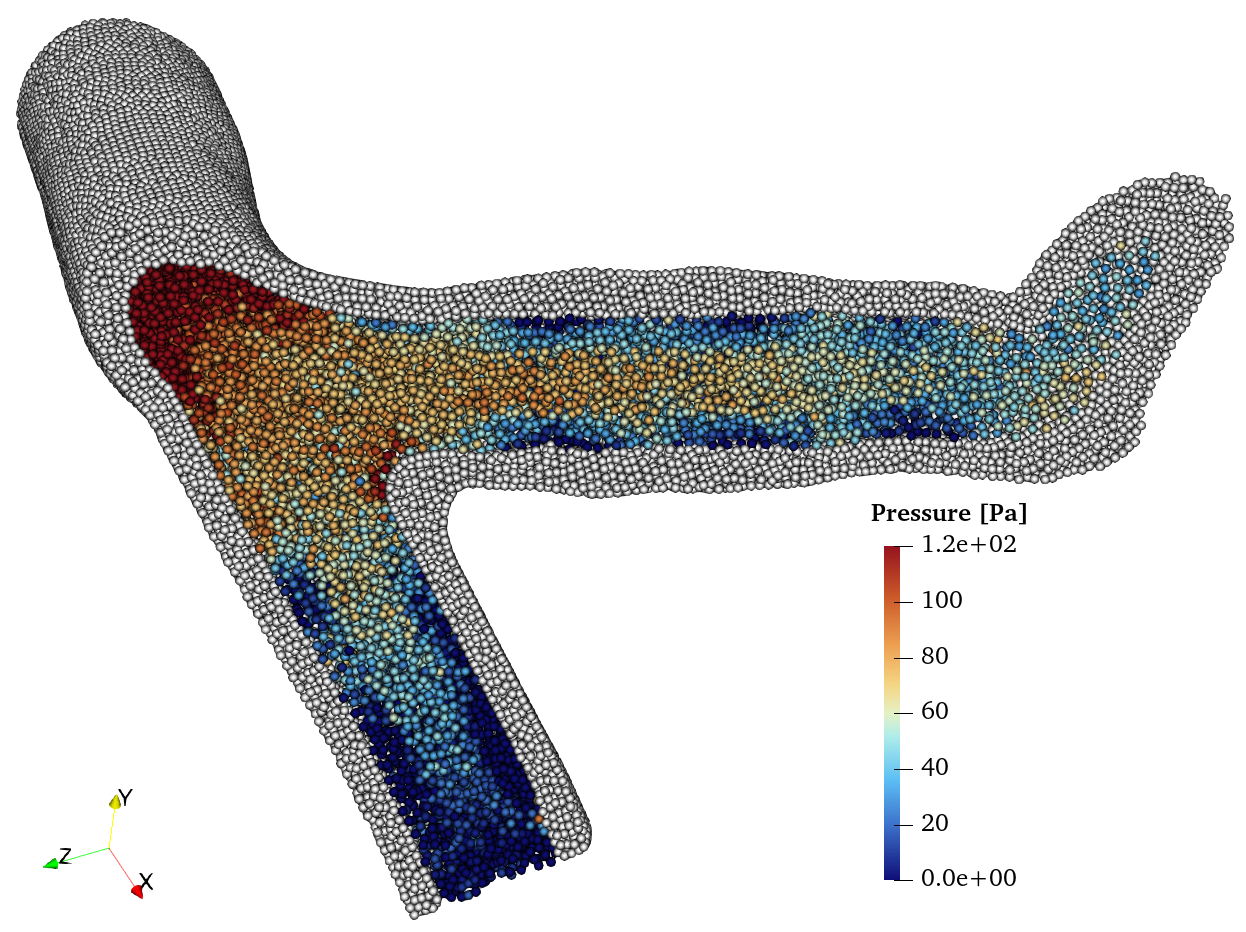}
        \vspace{4pt}
        \centering (b)
        \label{fig:26}
    \end{minipage}
    \caption{The cross-sectional view of the stenotic region compares the pressure distribution in the coronary artery before (left) and after (right) stent deployment. Red indicates high pressure, and blue represents low pressure. The labeled arrows mark the inlet and three outlet regions.}
    \label{fig:26}
\end{figure*}

\subsection*{Pressure at the Inlet and Outlets}
  
  \textbf{Before Stent Deployment (\textcolor{blue}{\hyperref[fig:26]{Fig.~26}} (a)):} Pressure is highest at the inlet (red) and declines along the main artery. Near the stenotic region, the gradient increases sharply, indicating strong flow obstruction and high resistance, with elevated pressure upstream and a rapid drop downstream. The large outlet exhibits relatively low pressure (dark blue), suggesting smooth outflow; the middle and small outlets are slightly higher but still low overall, implying no major pressure buildup despite altered distribution.

  \textbf{After Stent Deployment (\textcolor{blue}{\hyperref[fig:26]{Fig.~26}} (b)):} The overall pressure field is more uniform. The inlet remains relatively high, but the axial gradient is smoother with no abrupt drops. The reduced gradient across the treated site shows that local resistance is lowered, enabling easier passage and preventing excessive local peaks or sudden losses. The pressure at each outlet becomes more evenly distributed compared to the pre-stent condition, suggesting a more balanced redistribution of blood flow. This reduces localized high-pressure regions, leading to a more stable and steady blood flow.
  
\subsection*{Pressure Variation Across the Stenotic Region}
  
  \textbf{Before Stent Implantation:}The pressure before the stenosis (proximal-stenosis) is significantly higher than the distal-stenosis region, creating a sharp pressure drop across the stenotic segment. This sudden decrease in pressure can lead to hemodynamic instability, potentially increasing the risk of further vascular complications.

  \textbf{After Stent Implantation:} The cross-lesion pressure change is smoother, showing successful reduction of localized resistance, promoting a more stable hemodynamic environment and preventing excessive energy loss in the flow.
  
\subsection*{Improvement in Hemodynamic Stability}
  
  \textbf{Before Stent Implantation:} A high upstream pressure followed by a steep downstream drop, leads to an unfavorable flow condition that may contribute to increased shear stress on the vessel walls.

  \textbf{After Stent Implantation:} The pressure gradient is significantly reduced and the transition is smooth, stabilizing overall conditions, lowering the risk of endothelial damage and improving coronary perfusion.

\subsection*{Section Summary}
Overall, stent implantation reduces the excessive pressure gradient at the stenotic site, stabilizes the pressure field, and removes the abnormal high-speed jet there, yielding a more uniform velocity profile. The gentler pressure and velocity transitions cut unnecessary energy losses and turbulence, lower flow resistance in the treated region, and enhance perfusion efficiency. Together these changes improve blood-flow stability, restore more normal vascular function, and mitigate adverse loading on the vessel wall, helping to prevent further arterial narrowing.

\section{Conclusion}
This study develops an SPH-based workflow to study coronary blood flow and simulate stent implantation. The method combines a WCSPH fluid solver, stable boundary handling, and a solid model for the stent. Basic tests (start-up Poiseuille flow, a pressure-driven channel case, and a three-ring impact) showed good accuracy and stable behavior, so the setup is reliable before moving to coronary cases.

This study then applied it to a stenotic coronary bifurcation. After stent expansion, flow became more balanced across the branches, resistance across the lesion dropped, and FFR rose from \textbf{0.45} to \textbf{0.91}. Velocity, pressure, and wall-stress maps all showed clearer, smoother patterns, consistent with a reopened lumen.

Practically, this particle approach avoids meshing issues and works well with patient-oriented inputs (pre- and post-geometry, physiologic inlets and outlets). It delivers quantitative outputs: flow split, pressure gradients, WSS, and FFR that can support clinical planning and reduce reliance on operator experience in complex anatomy.

Several extensions can further strengthen this work. Blood rheology can be modeled with richer non-Newtonian laws, and vessel material behavior can incorporate layered structure, anisotropy, and disease-related remodeling. Model the balloon expansion explicitly, with a deformable balloon and contact with the stent and vessel, rather than approximating the balloon as a uniform pressure load. Map the effects of device settings (strut layout, thickness, balloon pressure, placement) and boundary specifications (inlet waveforms) using systematic sensitivity studies with uncertainty quantification. Report confidence bands for the key outputs. A fully patient-specific pipeline, from imaging through segmentation and geometry cleanup to simulation and automated reporting, will enhance clinical utility. Prospective validation against measurements (FFR, Doppler, and pressure-wire data) and multi-center datasets will support accuracy and generalization. Importantly, the framework can be extended to full fluid–structure interaction (FSI) by coupling blood flow with deformable vessel and stent mechanics, including contact and friction, to co-simulate expansion and recoil.

Overall, the SPH-based coronary stent simulation provides a practical path to evaluate pre- and post-implant hemodynamics and to analyze the deployment process. The approach resolves flow patterns, velocity fields, and vessel-wall stresses, offering actionable insights for intervention. With the deformable vessel behavior, and stent–artery contact, the same framework can guide stent design and implantation strategies to reduce the risk of in-stent restenosis. It can be further extended to personalized medicine by incorporating patient-specific imaging for precise preoperative simulation, individualized device selection, and implantation planning.

\bibliography{literature}

\end{document}